\documentclass[twocolumn, twocolappendix]{aastex631}
\usepackage{hyperref}
\usepackage{booktabs}
\usepackage{lineno}

\begin{document}

\title{Dust Evolution in Simulated Multi-phase Galactic Outflows}

\author[0000-0001-6325-9317]{Helena M. Richie}
\affiliation{Physics and Astronomy Department, University of Pittsburgh, 3941 O’Hara St, Pittsburgh, PA 15260 }

\author[0000-0001-9735-7484]{Evan E. Schneider}
\affiliation{Physics and Astronomy Department, University of Pittsburgh, 3941 O’Hara St, Pittsburgh, PA 15260 }

\begin{abstract}

We present the first large-scale, high-resolution simulations of dusty, star formation feedback-driven galactic outflows. Using the Cholla hydrodynamics code, we investigate dust sputtering in these environments for grains ranging in size from $1-0.001~{\mu\mathrm{m}}$. We compare results for two feedback models: one representative of low-redshift nuclear starburst galaxies and one similar to high-redshift main sequence galaxies. In general, our simulations show that multi-phase outflows are capable of safely transporting a vast majority of their dust to large distances ($\sim10~\textrm{kpc}$) from the disk. This work also shows that environmental shielding in cool gas clouds boosts dust survival rates significantly. The evolutionary path of dust depends strongly on grain size. Large grains ($a\geq0.1~{\mu\mathrm{m}}$) can be transported efficiently in all phases. Smaller grains, however, experience significant destruction in the hotter phases. $0.001~{\mu\mathrm{m}}$ grains in particular are quickly sputtered in all but the coolest gas, resulting in these grains strongly tracing the cool phase in outflows. These results may also indicate the importance of in-situ formation mechanisms, such as shattering, for the small dust grains and PAHs observed in emission throughout outflows in nearby galaxies. Surprisingly, we find that the hot phase dominates the transport of dust that survives to populate the CGM.

\end{abstract}


\section{Introduction} \label{sec:intro}

The galactic baryon cycle, which consists of galaxy-scale outflows and inflows, is central to galaxy evolution. The outflows driving the baryon cycle often originate from stellar feedback and can launch gas, metals, and dust from the interstellar medium (ISM) into the circumgalactic medium (CGM) and intergalactic medium (IGM). Key among these is dust, which provides a unique tracer for studying the baryon cycle due to its strong observational signature and interstellar origins. Because of these, the buildup of dust in regions beyond the ISM provides crucial insight into the baryon cycle that other tracers cannot. However, in order to use dust as a tool to study the baryon cycle, we must understand how dust evolves throughout this cycle.

Large-scale studies of dust-induced reddening in gravitationally lensed background quasars have revealed that dust is ubiquitous in the CGM \citep{Menard2010, McGee2010, Peek2015, McCleary2025}. In particular, \citet{Menard2010} detected reddening in millions of galaxy halos out to Mpc scales at $z\sim0.3$, suggesting that vast populations of extragalactic dust exist on huge scales. Based on these observations, the cosmic dust density, $\Omega_{\rm dust}$, of the halo at this redshift was estimated to be $\Omega_\mathrm{halo}^\mathrm{dust}\simeq2.1\times10^{-6}$, slightly higher than the estimated value for galactic disks, $\Omega_\mathrm{disk}^\mathrm{dust}\sim2\times10^{-6}$ \citep{Fukugita2004, Driver2007}. A wide variety of observational methods have corroborated the prevalence of dust in the CGM, including indirect measurements of dust absorption, extinction, and scattering in galaxy halos \citep{Zaritsky1994, Howk1997, Menard2012, HodgesKluck2014, Chen2025, Veilleux2025}, inferred dust masses through spectral energy distribution fitting \citep{Driver2018}, and direct observations of dust emission \citep{Engelbracht2006, Irwin2006, Contursi2013, Beirao2015, Levy2023, Bolatto2024, Chastenet2024}. These observations extend out to $z\sim5$, with $\Omega_\mathrm{dust}^\mathrm{halo}$ peaking at $z\sim1-1.5$ \citep{Chiang2025}.

Given its widespread presence and strong observational signature, dust has the potential to be a powerful tracer of the galactic baryon cycle. However, dust is not a static tracer, and our understanding of its evolution and impact on the baryon cycle is currently limited. In particular, the transport mechanism resulting in the large amounts of dust we observe in the CGM has remained relatively unexplored. Since $\Omega_\mathrm{halo}^\mathrm{dust}\sim\Omega_\mathrm{disk}^\mathrm{dust}$, and dust formation occurs predominantly within galaxies \citep{Dwek1980, Gehrz1989, Moseley1989} (although see \citet{Draine1990} for a discussion of possible extragalactic dust formation mechanisms), there must be a mechanism for the extremely efficient transport of dust from the ISM to the CGM. The galaxy-scale outflows that drive the galactic baryon cycle are an obvious candidate for this mechanism. However, their multi-phase nature complicates this proposed picture.

Observations have shown that galactic outflows are multi-phase, and are a common feature of rapidly star-forming galaxies. Down-the-barrel observations of blue-shifted absorption line spectra have revealed cool ($\sim10^4~\textrm{K}$) clouds moving at speeds exceeding the galactic escape velocity \citep{Martin1998, Rupke2005, Weiner2009, Rubin2010, Rubin2014}. These clouds are believed to be embedded in a hot, volume-filling wind originating from supernova feedback that breaks out of the disk and sweeps cool ISM material into the outflow \citep{Spitzer1956, Griffiths2000, Strickland2007, Lopez2020}. Theoretical work supports this picture, notably through large-scale simulations of galaxies with stellar feedback that can efficiently entrain cool clouds of ISM gas to form multi-phase outflows \citep{Kim2017, Fielding2018, Schneider2018, Schneider2020, Nguyen2022, Schneider2024}.

While dust-rich ISM clouds entrained in multi-phase outflows may transport dust out of galaxies, it is not immediately clear that dust can survive in these environments. Dust grains are not static objects, but can evolve in size, composition, and distribution throughout their lifecycle depending on their environment. In particular, in hot gas and shocks, such as in supernovae, dust grains can be destroyed through sputtering \citep{Draine1979a, Draine1979b, Jones1994}. We currently have a limited understanding of how these mechanisms may impact dust on its journey to the CGM in galactic outflows. Because outflows are driven by hot gas and contain turbulent cool gas \citep{Qu2022, Chen2023, Abruzzo2024}, it seems likely that sputtering and shattering could significantly affect their dust populations. An understanding of how dust evolves on its way to the CGM is thus crucial to interpreting observations such as \citet{Menard2010}, which rely on assumptions about the dust grain size distribution and composition, among other things \citep{Mathis1977, Weingartner2001, Hensley2023}.

Many simulations have explored the evolution of dust over cosmic time using large-scale and cosmological simulations (e.g. \citealt{Zu2011, Bekki2015, Popping2017, Aoyama2018, Li2019, Choban2022}). However, relatively few simulations have explored the evolution of dust as it moves through outflows. So far, these have primarily consisted of ``cloud-crushing" simulations, where individual cool, dusty clouds are accelerated by a hot background wind (\citealt{Farber2022, Chen2024, Richie2024}; although see also \citealt{Kannan2021}). These works have focused on investigating the ability of cool clouds to shield dust from destruction in the hot wind, as dust lifetimes in the cool phase can be of order $\sim$\textrm{Gyr} ($\gtrsim$thousands of times longer than in the hot phase). The ``cloud-crushing" simulation setup has been studied in detail in the context of cloud entrainment in galactic outflows \citep{Klein1994, Cooper2009, Schneider2017, Gronke2018, Abruzzo2022a}. These studies primarily focus on understanding the ability of clouds to survive as they are launched into outflows, finding that some clouds can survive long-term and even grow, while others can be destroyed \citep{Sparre2020, Abruzzo2023}. \citet{Richie2024} investigated dust evolution due to sputtering in a variety of cloud evolutionary scenarios and found, overwhelmingly, that outflows could transport large amounts of dust to the CGM. However, that work showed that dust survival is more sensitive to grain size than to cloud shielding, with a majority of larger grains surviving even in the hot phase (due to the wind's adiabatic expansion and subsequent rapid drop in density and temperature), while small grains experience significant sputtering regardless of cloud shielding.

These simulations have helped us make strides in our understanding of dust's role in the baryon cycle by providing an explanation for the huge populations of extragalactic dust observed in \citet{Menard2010}. However, a number of open questions remain, including how the grain size distribution and overall dust content evolve in global outflow environments, as well as which dust evolution mechanisms are relevant in these regions. Particularly, although \citet{Richie2024} showed that a majority ($\gtrsim60\%$) of the large dust grains launched into outflows can survive to populate the CGM, the \citet{Menard2010} observations imply that the halo and disk dust contents are comparable to one another. It is possible that environmental shielding in global galactic outflows may enable enhanced dust survival compared to idealized cloud-wind simulations. Similarly, the efficient sputtering of small grains measured in \citet{Richie2024} disagrees with the abundance of small dust grains and polycyclic aromatic hydrocarbons (PAHs) observed far from the disk mid-plane in, e.g., nearby starburst galaxy M82 \citep{Bolatto2024}.

To address these questions, in this paper, we present global simulations of two star-forming galaxies driving dusty galactic outflows. We model the evolution of dust grain sizes ranging from $1-0.001~{\mu\textrm{m}}$ using the sputtering model developed in \citet{Richie2024}. These simulations provide us with the first high-resolution ($\Delta\textrm{x}<5~\textrm{pc}$), global maps of dust distribution in large-scale multi-phase galactic outflows, as well as insight into dust evolution as it moves to the CGM. This paper is organized as follows: in Section~\ref{sec:simulations}, we describe our dust model and the galaxy models used in these simulations. In Section~\ref{sec:results}, we present our results, including an overview of the time-series evolution, a detailed look at the dust distribution in the $30~\textrm{Myr}$ snapshot, and overall dust survival fractions. In Section~\ref{sec:discussion}, we discuss the implications of these results for large- and small-scale observations of extragalactic dust, the effects of global outflow environments on sputtering, and the future work needed to fully simulate dust evolution in outflows. In Section~\ref{sec:conclusions}, we summarize our main takeaways.

\section{Simulations} \label{sec:simulations}

In this Section, we briefly describe our initial conditions and stellar feedback model. All simulations were run with the Cholla hydrodynamics code \citep{Schneider2015}, using the HLLC Riemann solver \citep{Toro1994, Batten1997}, a piecewise-parabolic interface reconstruction \citep{Fryxell2000}, and the
second-order unsplit Van Leer integrator \citep{Stone2009}. Each simulation was carried out at a constant resolution of $\Delta x \approx4.9~\text{pc}$ in a $L_x\times L_y\times L_z = 10\times 10\times 20~\text{kpc}^3$ box on a Cartesian grid of $2048\times2048\times4096$ cells. We employ diode boundary conditions at all boundaries, which allow material to flow out of the volume but prevent inflow. Gas cooling is modeled using a parabolic fit to a collisionally ionized equilibrium (CIE) Cloudy cooling curve \citep{Ferland2013, Schneider2017}, with zero cooling below $10^4~\text{K}$, assuming a balance between cooling and heating from the photoionizing UV background.\footnote{Limiting cooling to $10^4~{\rm K}$ will impact our simulation's ability to form high-density, cool gas below the cooling floor. However, similar high-resolution simulations that allow gas to cool to lower temperatures find that high-density gas in outflows primarily exists in fountain flows near the disk (e.g., \citealt{Kim2020}). Therefore, our choice of cooling curve is unlikely to significantly impact our results for dust evolution in the outflow and CGM dust loading.}

The disk and halo model and initial conditions are described in detail in \citet{Schneider2018}, but we briefly describe them here. The initial conditions for the stellar and gas distributions are modeled after the well-studied M82 galaxy, a nearby starburst with a bi-conical outflow. Initially, we place an exponential disk of $10^4~\textrm{K}$ gas in the center of the box, with a mass of $M_\mathrm{disk,gas}=2.5\times10^9~M_\odot$ and scale radius of $R_\mathrm{disk,gas}=1.6~\textrm{kpc}$ \cite{Greco2012}. The disk is initially in hydrostatic and rotational equilibrium with a static gravitational potential, following a Miyamoto–Nagai \citep{Miyamoto1975} and a Navarro--Frenk--White \citep{Navarro1996} profile for the stellar and dark matter components, respectively. The stellar disk has a mass of $M_\mathrm{disk,stars}=10^{10}~M_\odot$, scale radius of $R_\mathrm{disk,stars}=0.8~\textrm{kpc}$ \citep{Mayya2008} and the dark matter profile has a mass of $M_\mathrm{halo}=5\times10^{10}~M_\odot$, scale radius of $R_\mathrm{halo}=5.3~\textrm{kpc}$, and concentration of $c=10$. A hot halo surrounding the galaxy is initialized with $n\approx10^{-3}~\text{cm}^{-3}$ and $T\approx2\times10^6~\text{K}$, but is blown out of the box early on in the simulation.

Stellar feedback is generated collectively by individual star clusters distributed throughout the disk that inject mass and thermal energy into their surrounding medium. This feedback model is described in detail in \citet{Schneider2024}. We use an updated cluster mass function (CMF) slope ($\alpha=1.9$; \citealt{Levy2024}) and a maximum cluster mass of $M_\mathrm{cl,max}=2.5\times10^6~M_\odot$ \citep{McCrady2007, Levy2024}. To limit computational expense, we set the minimum cluster mass to $M_\mathrm{cl,min}=10^4~M_\odot$. Clusters within this mass range are periodically placed into the disk at the star formation rate (SFR) with masses determined by a probability density function proportional to $M^{-\alpha}$. The location of the cluster is set by a distribution function

\begin{equation}
N_\mathrm{cl}\propto R\,e^{-R/1.0~\textrm{kpc}} \label{eq:cluster_dist}
\end{equation}

\noindent where $R$ is the cluster distribution scale radius. We run two fiducial simulations, representing star-forming galaxies at low and high redshifts, in which we vary the SFR and the cluster distribution scale radius. The low-redshift-analogue \texttt{nuclear-burst} model contains centrally concentrated star clusters ($R=0.3~\textrm{kpc}$), similar to the distribution observed in M82, and a SFR of $5~M_\odot\,\textrm{yr}^{-1}$. The \texttt{high-z} model is more in line with higher-redshift galaxies, with a disk-wide distribution of star clusters ($R=0.8~\textrm{kpc}$), and a SFR consistent with galaxies of this mass on the star-forming main sequence at $z\sim2$ \citep{Whitaker2012}.

We employ Cholla's scalar-based dust model (see \citealt{Richie2024} for full details) to simulate the evolution of dust density throughout the grid. Initially, all dust grain sizes are evenly distributed throughout the disk with an equal dust-to-gas mass ratio (DGR) of 1:100, similar to the DGR of the Milky Way \citep{Draine2011}. We note that our analysis primarily focuses on the relative evolution of gas and dust, so the choice of DGR does not have an important effect on our results. The total dust density in each cell evolves subject to sputtering

\begin{equation}
    \frac{\textrm{d}\rho_\mathrm{dust}}{\textrm{d}t}=-\frac{\rho_\mathrm{dust}}{t_\mathrm{sp}/3}.
\end{equation}

\noindent Where $\rho_\mathrm{dust}$ is the cell's dust density and $t_\mathrm{sp}$ is the sputtering time,

\begin{equation}
t_\mathrm{sp}=0.17~\textrm{Gyr}\Big(\frac{a}{0.1~\mu\text{m}}\Big)\Big(\frac{10^{-27}~\textrm{g}\,\textrm{cm}^{-3}}{\rho_\mathrm{gas}}\Big) \Big[\Big(\frac{10^{6.3}~\textrm{K}}{T_\mathrm{gas}}\Big)^\omega+1\Big].
\label{eq:sput-timescale}
\end{equation}

\noindent  Here, $a$ is the radius of a (spherical) dust grain and $\rho_\mathrm{gas}$ and $T_\mathrm{gas}$ are the local gas density and temperature, respectively.\footnote{Note that Equation~\ref{eq:sput-timescale} does not consider the finite length of grain material, which has an important effect on the sputtering yield of small grains within hot ionized gas \citep{Kirchschlager2019}.} At high temperatures (i.e. above $10^{6.3}~\textrm{K}$), $t_\mathrm{sp}$ becomes roughly constant. This temperature is set by the constant $\omega=2.5$.

We include several dust scalar fields to simultaneously, but independently, model a range of grain sizes. In both of our fiducial simulations, we include four grain sizes: $a=1$, $0.1$, $0.01$, and $0.001~{\mu\textrm{m}}$, chosen to span the Milky Way grain size distribution \citep{Weingartner2001}. 

\section{Results} \label{sec:results}

\begin{figure*}
\centering
\includegraphics[width=0.9\textwidth]{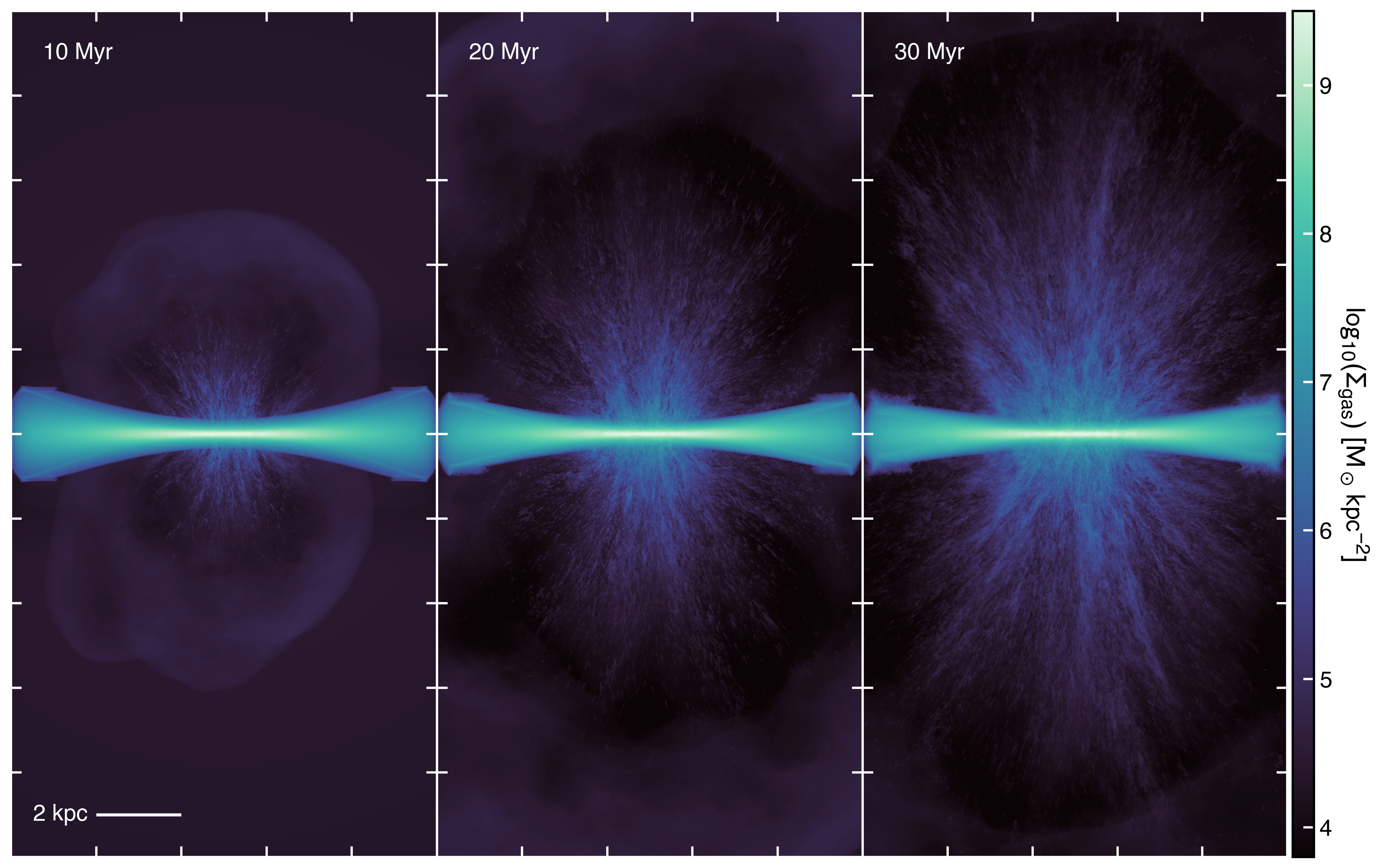}

\vspace{0.5cm}

\includegraphics[width=0.9\textwidth]{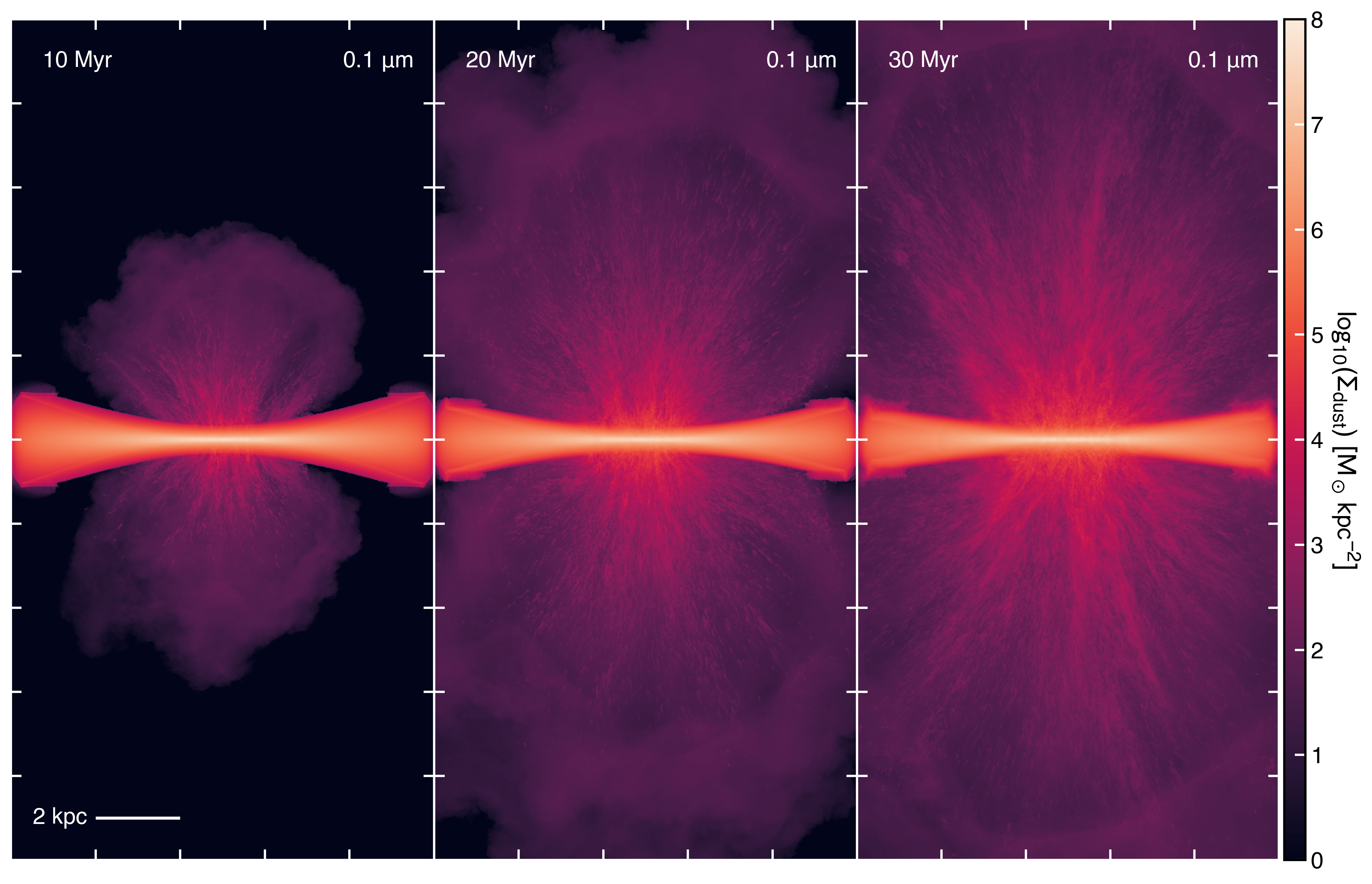}


\caption{Projections of the gas surface density ($\Sigma_\mathrm{gas}$, top) and $a=0.1~{\mu\mathrm{m}}$ dust surface density ($\Sigma_\mathrm{dust}$, bottom) through the $xz$-plane for the \texttt{nuclear-burst} simulation at 10, 20, and $30~\textrm{Myr}$. Each panel shows the full extent of the $xz$-plane, $10\times20~\textrm{kpc}^2$. An animated version of this figure is available in the online article.}
\label{fig:m82_proj_evolve}
\end{figure*}

\begin{figure*}
\centering
\includegraphics[width=0.9\textwidth]{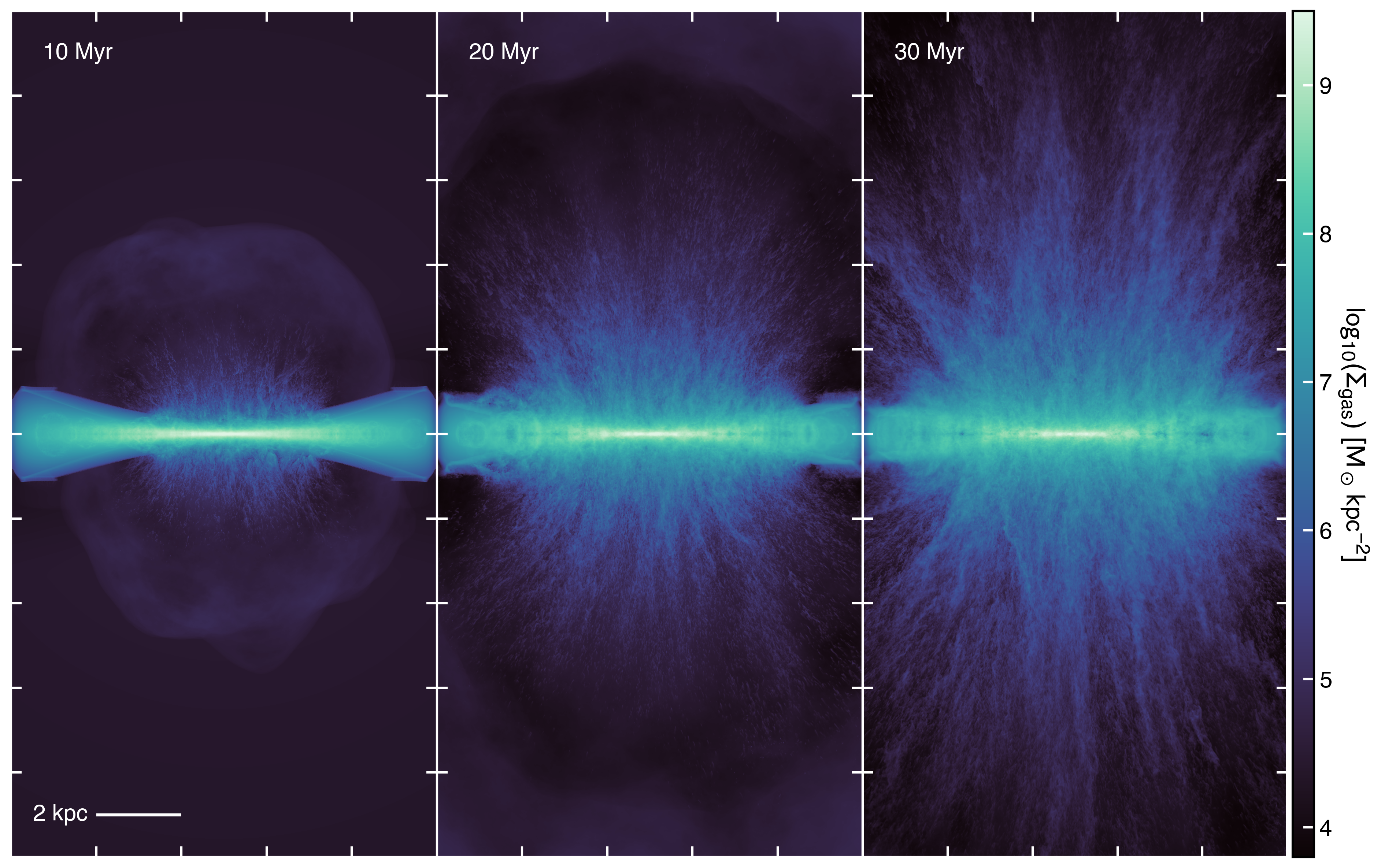}

\vspace{0.5cm}

\includegraphics[width=0.9\textwidth]{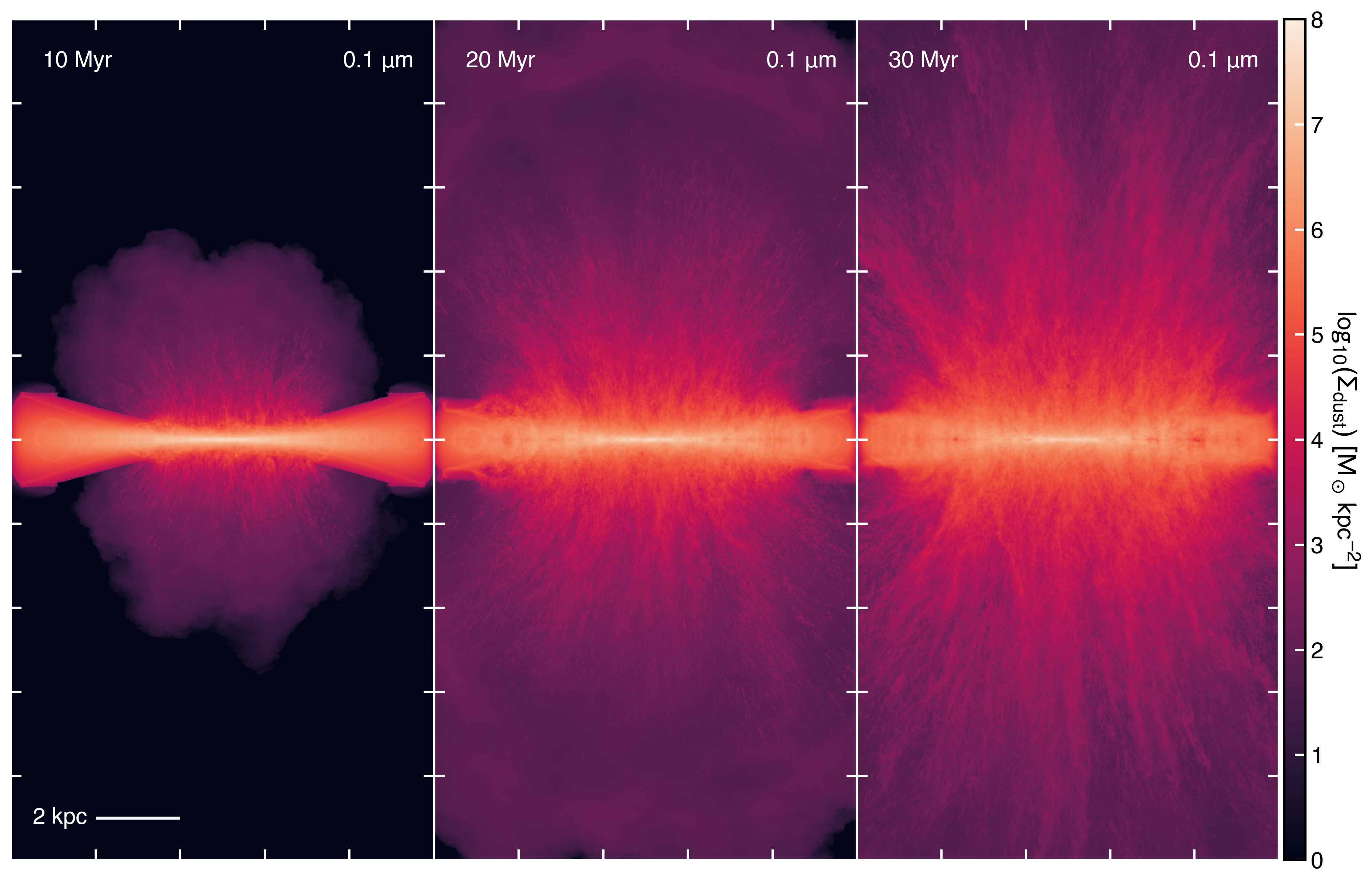}
\caption{Same as Figure~\ref{fig:m82_proj_evolve}, but for the \texttt{high-z} simulation. An animated version of this figure is available in the online article.}
\label{fig:highz_proj_evolve}
\end{figure*}

\begin{figure*}[p]
\centering
\includegraphics[width=0.86\textwidth]{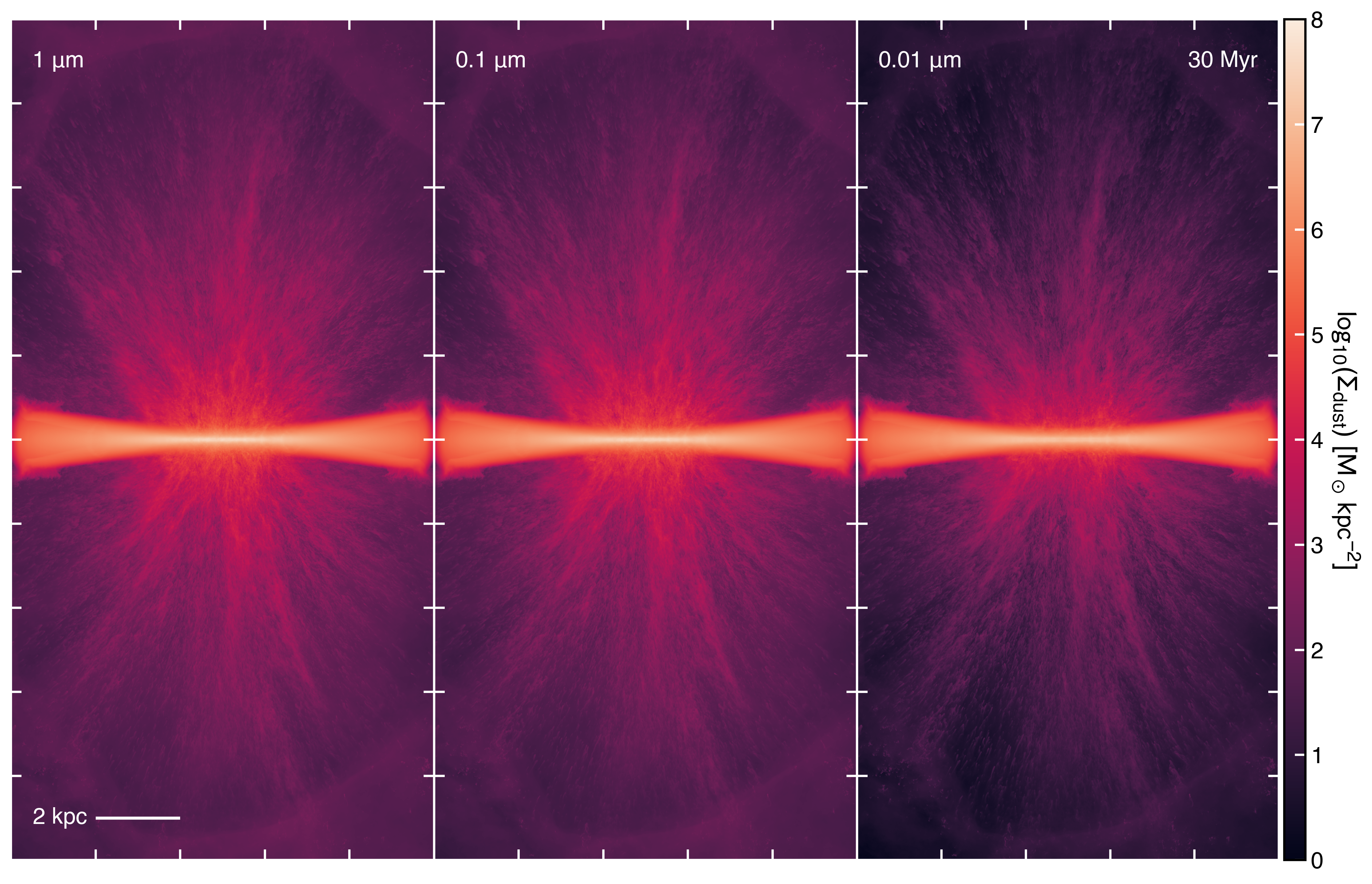}

\vspace{0.5cm}

\centering
\includegraphics[width=0.86\textwidth]{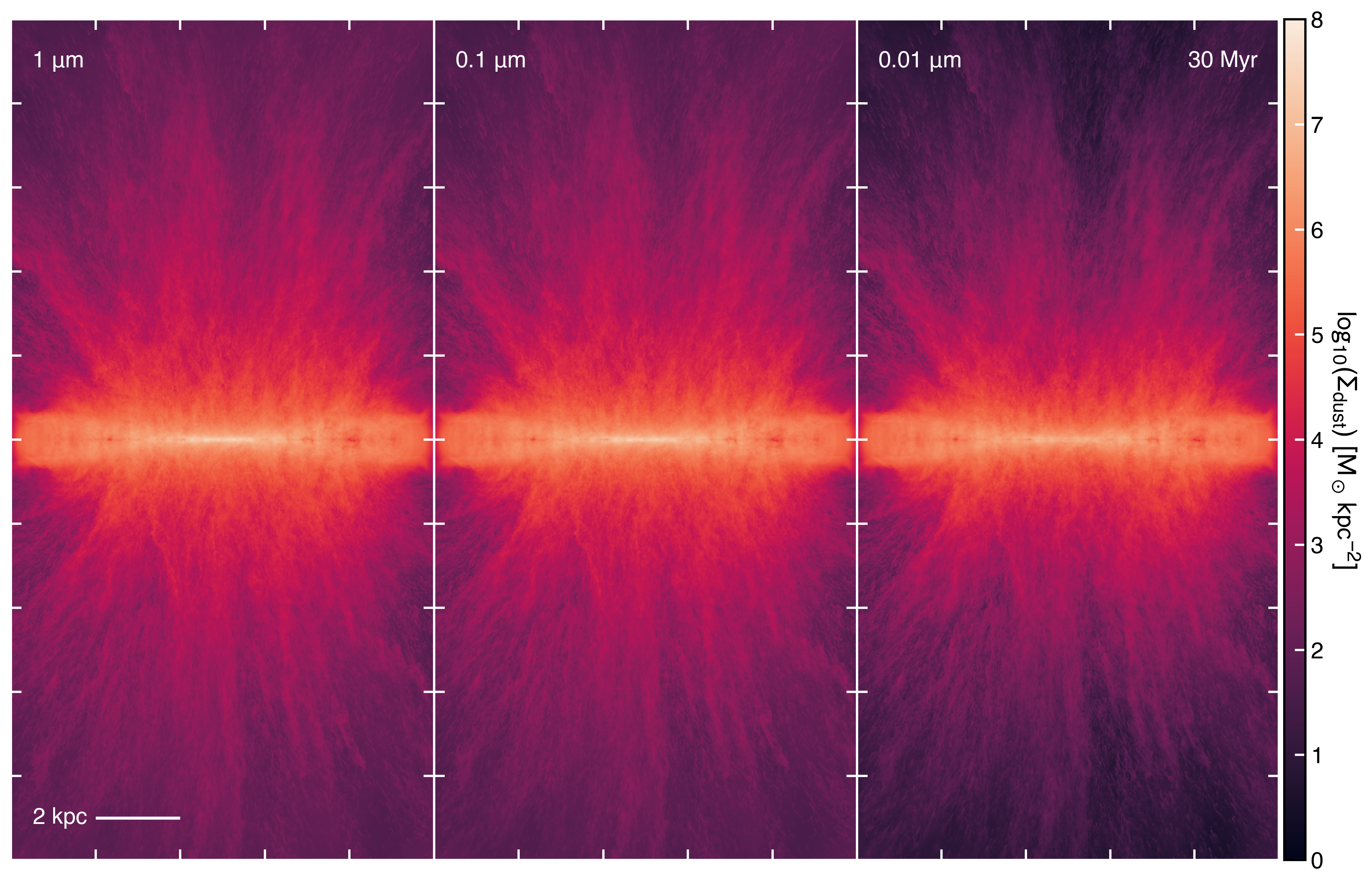}

\caption{Dust surface density projections for $1$, $0.1$, and $0.01~{\mu\textrm{m}}$ radius dust grains at $30~\textrm{Myr}$. The top and bottom row shows the \texttt{nuclear-burst} and \texttt{high-z} simulations, respectively. An animated version of this figure is available in the online article.} 
\label{fig:dust_projections}
\end{figure*}

We first give a qualitative overview of each simulation and its time evolution in Section~\ref{subsec:overview}. Then, we focus on simulation snapshots at $30~\textrm{Myr}$ and describe their gas and dust outflow properties (Sections~\ref{subsec:radial_profiles} and \ref{subsec:dust_properties}, respectively), dust sputtering in the outflow (Section~\ref{subsec:sputtering}), outflow dust-to-gas ratios (Section \ref{subsec:dgr}), dust outflow rates (Section~\ref{subsec:outflow_rate}), and dust survival fractions (Section \ref{subsec:survival_fracs}). We then turn to snapshots at $15~\textrm{Myr}$ to discuss the evolution of the smallest dust grains at early times (Section~\ref{subsec:small_dust}).

\subsection{Qualitative simulation overview} \label{subsec:overview}

Each simulation begins with a smooth distribution of gas and dust in the disk, surrounded by a hot, static, and dust-free halo. Star clusters are placed throughout the disk according to Equation~\ref{eq:cluster_dist} and begin to inject dust-free mass and energy into the surrounding ISM at a rate determined by the target SFR of the simulation. Stellar feedback then drives a wind of hot, thermalized ejecta that breaks out of the disk and into the halo. Over time, this feedback also generates an outflow of cool, dense, and dusty gas from the disk.

Figure~\ref{fig:m82_proj_evolve} shows projections of the gas ($\Sigma_\mathrm{gas}$, top) and $a=0.1~{\mu\textrm{m}}$ dust ($\Sigma_\mathrm{dust}$, bottom) surface density in the \texttt{nuclear-burst} simulation at $10~\textrm{Myr}$, $20~\textrm{Myr}$, and $30~\textrm{Myr}$, and Figure~\ref{fig:highz_proj_evolve} shows the same projections for the \texttt{high-z} simulation. At $10~\textrm{Myr}$, the feedback-driven wind has only reached a few kpc from the disk in both simulations. The leading edge of the wind-driven bubble is shown by the forward shock, which appears as a sharp boundary in $\Sigma_\mathrm{gas}$ furthest from the galaxy. Another $\Sigma_\mathrm{gas}$ boundary can be seen in the outflow at the location of the reverse shock, which propagates back into the outflow after the initial collision between the wind and the halo. Looking at $\Sigma_\mathrm{dust}$, the edge of the dust distribution roughly marks the contact discontinuity between the two shock fronts, where the wind and halo initially collide. Within this boundary, dust fills the entire outflow region. At early times, the \texttt{nuclear-burst} and \texttt{high-z} outflows have somewhat different shapes, with the \texttt{nuclear-burst} outflow appearing more biconical and the \texttt{high-z} outflow more spherical. Most of the outflow at this point consists of the hot, diffuse wind. However, at the base of the outflow, filaments of dense gas and dust can be seen beginning to move out of the disk.

At $20~\textrm{Myr}$, feedback continues to drive the dusty wind outward, and it has expanded to fill most of the simulation box. The contact discontinuity and reverse shocks are both still at least partially visible at $20~\textrm{Myr}$, but more so for \texttt{nuclear-burst} since its wind expands more slowly than the \texttt{high-z} wind. The dense, dusty filaments are still being accelerated (by mixing with the hot wind) and thus have only traveled a few kpc further than in the $10~\textrm{Myr}$ snapshot. Within these filaments, there is a noticeable drop in both $\Sigma_\mathrm{gas}$ and $\Sigma_\mathrm{dust}$ with distance from the galaxy.

Now, we limit our focus to the $30~\textrm{Myr}$ snapshot, which allows us to compare directly with results from previous CGOLS studies that also focus on 30 Myr snapshots \citep{Schneider2020, Schneider2024}. At this point, the dense filaments of gas and dust have expanded nearly to the edge of the volume. The decrease in filament density with distance from the disk is still visible in both $\Sigma_\mathrm{gas}$ and $\Sigma_\mathrm{dust}$. The reverse shock is no longer present in the \texttt{high-z} simulation, and it has almost left the box in the \texttt{nuclear-burst} simulation as the wind continues to expand.

Comparing Figure \ref{fig:m82_proj_evolve} and Figure \ref{fig:highz_proj_evolve} at $30~\textrm{Myr}$, the most prominent difference is the distribution of gas and dust in the outflow, caused by the simulations' differing star cluster distributions and SFR. The \texttt{nuclear-burst} model's centrally concentrated clusters drive an outflow primarily in the central region of the disk that expands outward radially, resulting in a cone-like morphology. The more distributed clusters in the \texttt{high-z} model lead to an outflow that is widespread throughout the disk with a more vertical geometry. The \texttt{high-z} model's higher SFR also drives gas out of the disk faster than in the \texttt{nuclear-burst} model, leading to a more developed outflow in the \texttt{high-z} simulation. Evidence for this can be seen, for example, by the absence of the reverse shock in the \texttt{high-z} simulation at $30~\textrm{Myr}$, which is no longer visible at this point. Furthermore, despite a difference of only a factor of 4 in their SFRs, the \texttt{high-z} galaxy has about 15 times more cool gas in its outflow than \texttt{nuclear-burst} at this time.

In Figure~\ref{fig:dust_projections}, we illustrate how the dust morphology depends on grain size. Here we show $\Sigma_\mathrm{dust}$ for 1, 0.1, and $0.01~{\mu\textrm{m}}$ radius dust grains at $30~\textrm{Myr}$ in the \texttt{nuclear-burst} (top) and \texttt{high-z} (bottom) models. Differences in the dust distribution between grain sizes arise because of their differing sputtering timescales. For the two largest grain sizes ($1$ and $0.1~{\mu\mathrm{m}}$), the volume in both simulations is completely filled with dust. It is densest in the dense gas that corresponds to the cool phase, but it also permeates the hotter volume-filling phase. The $0.01~{\mu\mathrm{m}}$ grains are similarly dense in the cool phase but exhibit a higher level of depletion in regions of diffuse gas, which corresponds to the hot phase. The sputtering time of $0.01~{\mu\textrm{m}}$ grains in the hot wind is of order $\sim1-10~\textrm{Myr}$, explaining this behavior. Figure~\ref{fig:dust_projections} also shows that the effects of sputtering increase with distance from the galaxy as dust spends more time in the outflow.

\subsection{Outflow Gas Properties} \label{subsec:radial_profiles}

\begin{figure*}
\centering
\includegraphics[width=0.32\textwidth]{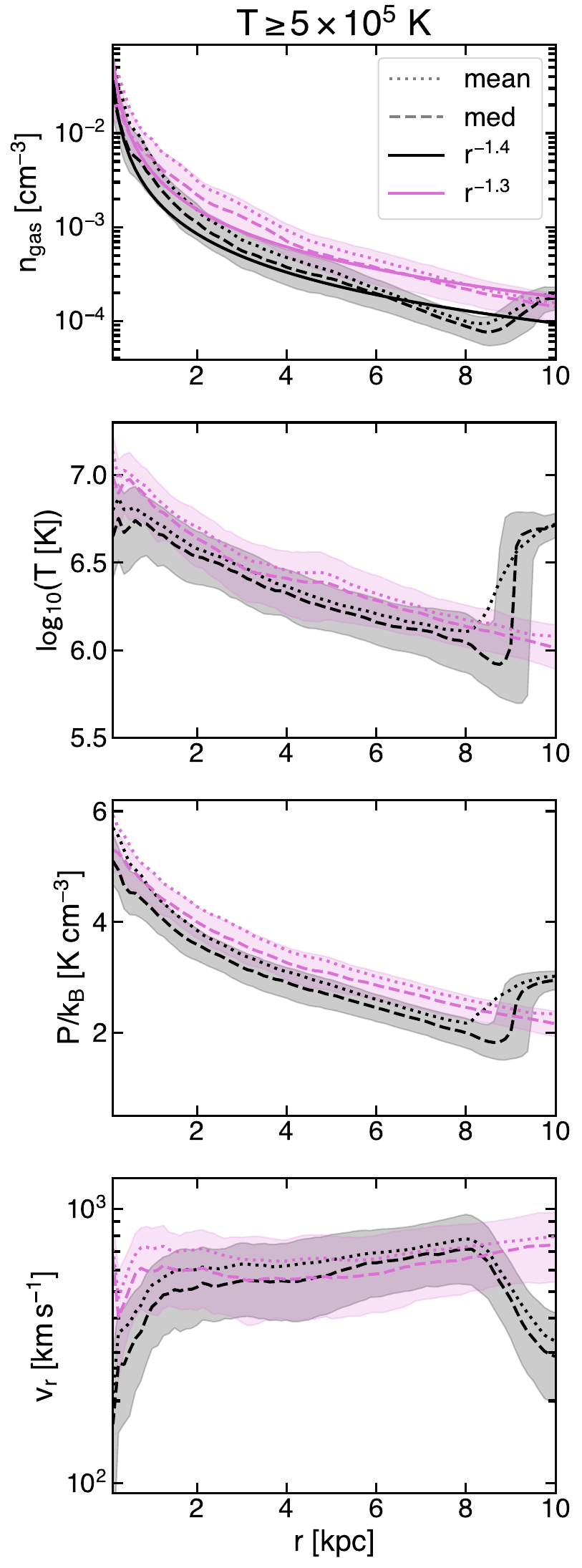}
\includegraphics[width=0.32\textwidth]{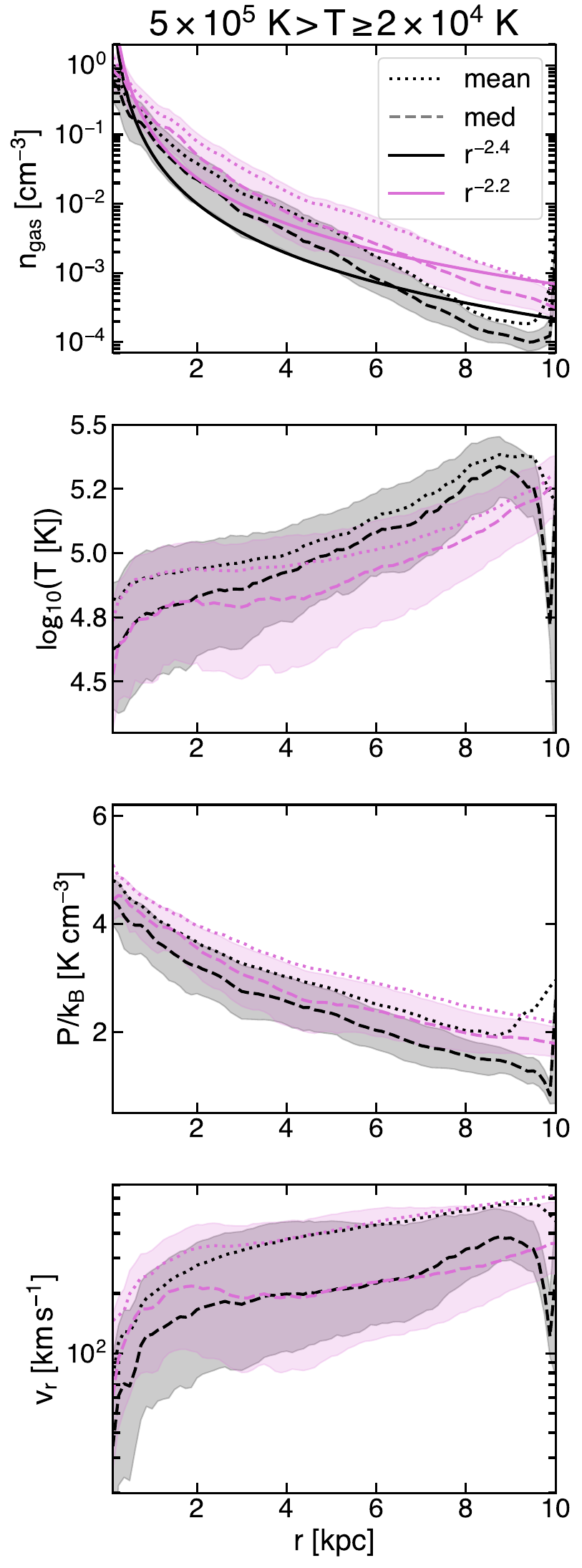}
\includegraphics[width=0.32\textwidth]{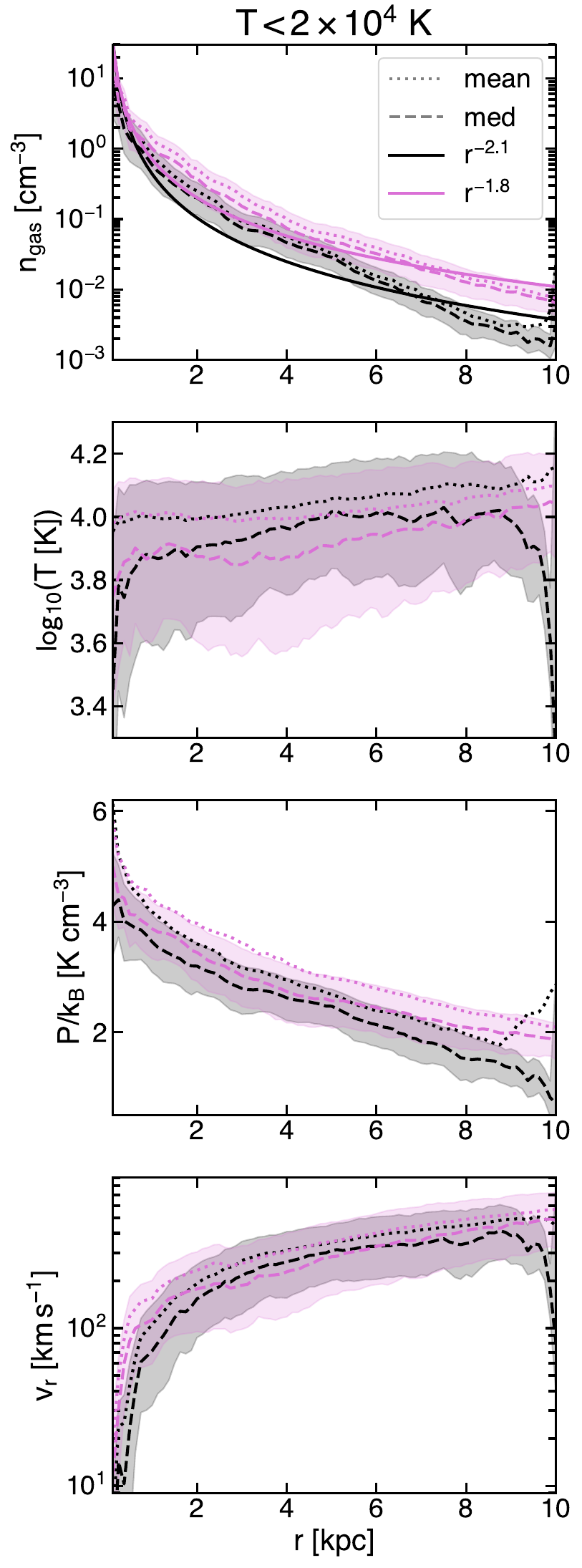}
\caption{Radial profiles for the hot (left), intermediate (middle), and cool (right) phases of the \texttt{nuclear-burst} (black) and \texttt{high-z} (pink) simulations, measured using $30^\circ$ cones opening away from the disk mid-plane. From top to bottom, the panels show gas number density, radial velocity, pressure, and temperature (all density-weighted). The dotted and dashed lines show the mean and median of each quantity, respectively, and the shaded regions show the 25th and 75th percentile range. The solid lines show power law fits to the median density profiles.}
\label{fig:radial_profiles}
\end{figure*}

\begin{figure*}
\centering
\includegraphics[width=0.7\textwidth]{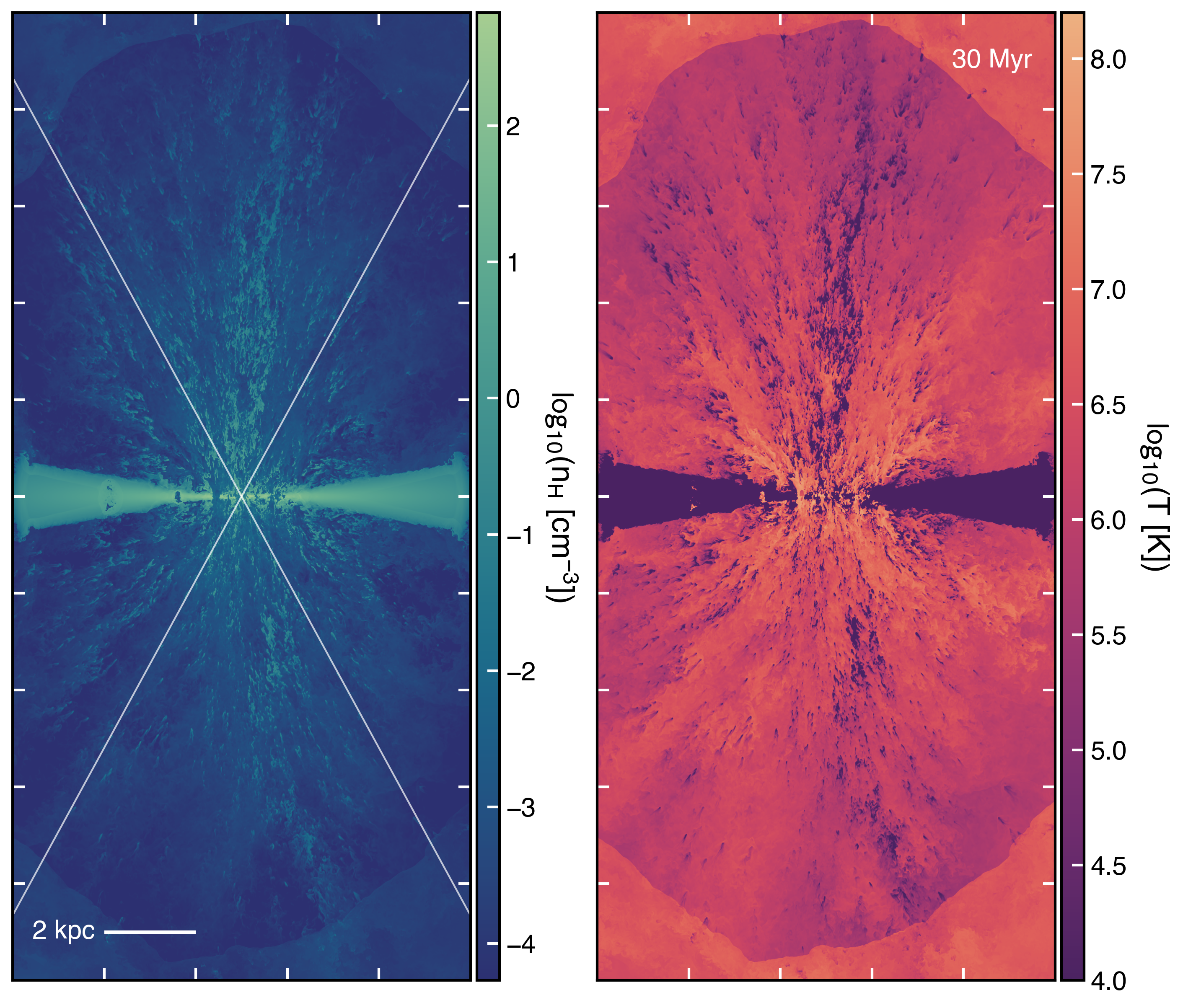}

\vspace{0.5cm}

\includegraphics[width=0.7\textwidth]{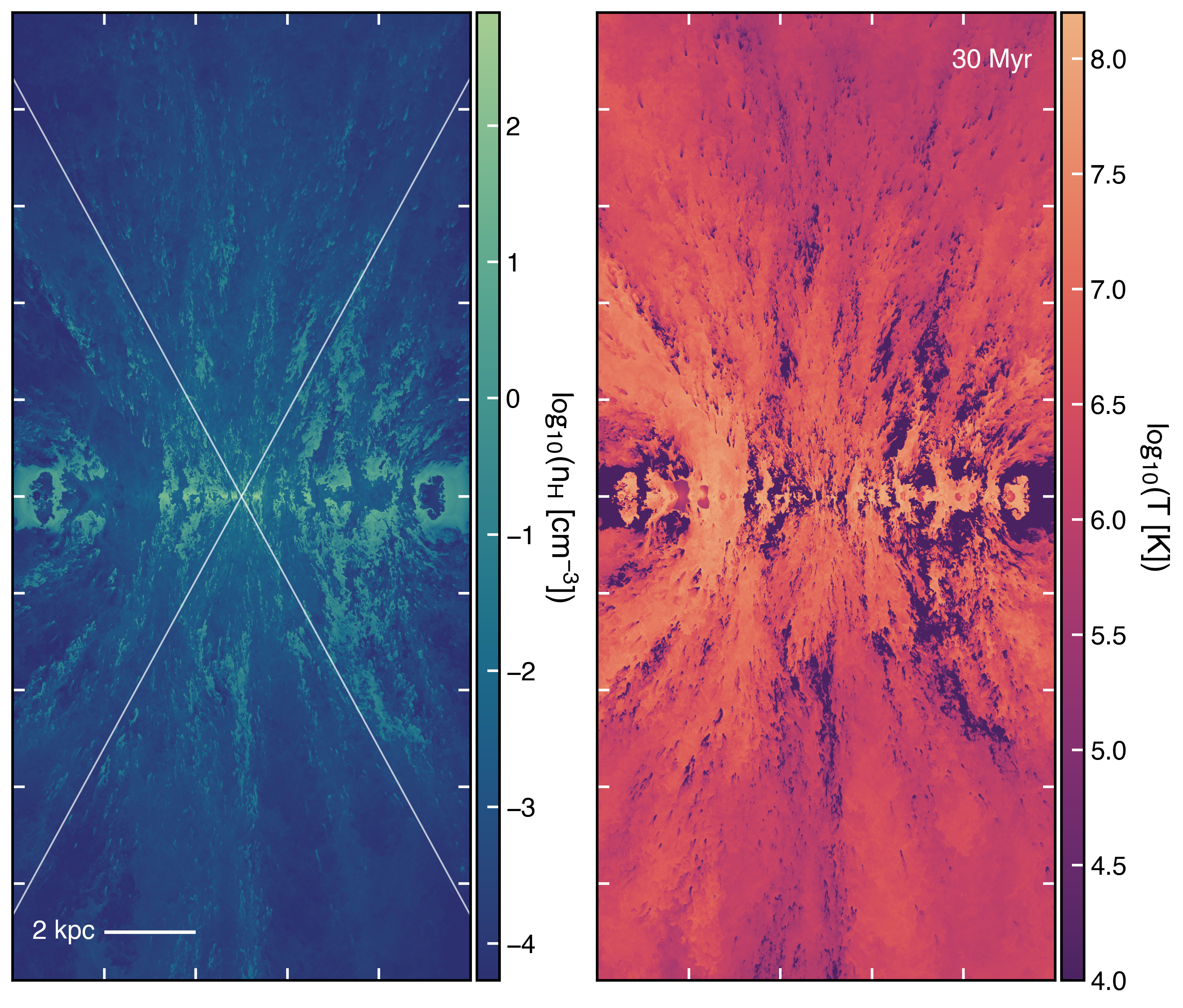}
\caption{Density (left) and temperature (right) slices through the center of the $xz$-plane of the \texttt{nuclear-burst} (top) and \texttt{high-z} (bottom) simulations at 30 Myr. The solid diagonal lines in the left panels show the $30^\circ$ cones used to measure the gas and dust radial profiles described in Section~\ref{subsec:radial_profiles}.}
\label{fig:gas_slices}
\end{figure*}

The evolution of dust depends critically on gas thermal and dynamical properties. To better understand dust's environment in the outflow, we measured radial profiles of the outflow gas properties of the $30~\textrm{Myr}$ simulation snapshots. Radial profiles for similar simulations have previously been discussed at length in \citet{Schneider2020} and \citet{Schneider2024}; further details can be found in these references. We used $30^\circ$ cones placed above and below the disk, opening away from the disk mid-plane into the outflow to measure various gas properties for the hot ($T\geq5\times10^5~\textrm{K}$), intermediate ($5\times10^5~\textrm{K}>T\geq2\times10^4~\textrm{K}$), and cool ($T<2\times10^4~\textrm{K}$) gas phases, following the phase definitions used in \citet{Kim2017} and \citet{Schneider2020}. The intermediate phase brackets the peak of the cooling curve and results in short cooling times (i.e. $t_{\rm cool}\sim10^5~{\rm yr}$) compared to the longer-lived hot and cool phases. We measured the gas density, radial velocity, pressure, and temperature within these cones using radial bins of width $r=0.125~\textrm{kpc}$. Figure~\ref{fig:radial_profiles} shows the resulting density-weighted average (dotted lines), median (dashed lines), and $25-75$ percentile range (shaded region) of these quantities as a function of $r$. The solid lines show a power law fit to each phase's median density profile. The \texttt{nuclear-burst} model is shown in black and the \texttt{high-z} model in pink. In Figure~\ref{fig:gas_slices}, we also show the gas density and temperature in slices along the center of the $xz-$plane. The solid diagonal lines here show the opening angle of the cone used to measure the radial profiles.

Characteristic behaviors of each phase are apparent in these profiles. First, the hot phase density and temperature fall by roughly 2-3 dex by $r=10~\textrm{kpc}$. This is caused by the outward expansion of the wind from the disk into the surrounding medium, though we note that the resulting density profiles are shallower than the $r^{-2}$ shape predicted by \citet{Chevalier1985} for the hot wind. The cool phase density also drops by several orders of magnitude as the clouds expand into the outflow, roughly following an $r^{-2}$ density profile (see power law fits). The cool phase temperature throughout the outflow remains slightly below $\sim10^4~\textrm{K}$, the temperature at which the cooling rate goes to zero, although cool gas can exist below this temperature if the clouds expand adiabatically. Similar to the hot and cool phases, the intermediate-phase density falls by several orders of magnitude throughout the outflow. The intermediate-phase temperature is, by definition, intermediate between the hot and cool phases, and increases by nearly an order of magnitude by $r=10~\textrm{kpc}$. The cool and intermediate phases are in rough pressure equilibrium with one another, but are slightly underpressurized compared to the hot phase.

The hot wind speed is relatively flat as a function of $r$, reaching $v_\mathrm{r}\sim800~{\textrm{km}\,\textrm{s}^{-1}}$ by $r=10~\textrm{kpc}$ in both simulations. The cool phase, on the other hand, accelerates from its initial $v_\mathrm{r}$ of zero by gaining momentum through mixing with the wind (i.e., through the cloud acceleration mechanisms described in, e.g., \citealt{Gronke2018}; \citealt{Schneider2020}; \citealt{Richie2024}). The velocity of the cool phase in both simulations approaches $v_\mathrm{r}\sim500~{\textrm{km}\,\textrm{s}^{-1}}$. Overall, there is a fairly smooth gradient in velocity between phases, mediated by mixing in the intermediate phase.

Differences between the gas properties of the two models can be seen by comparing the radial profiles. These discrepancies arise primarily due to their differing SFRs. In particular, the \texttt{high-z} model's 4x larger SFR leads to an overall denser outflow and hotter wind. This can be seen in both the radial profiles and the density and temperature slices in Figure~\ref{fig:gas_slices}. In the \texttt{high-z} slice, the entire region adjacent to the disk appears considerably hotter and denser than in the \texttt{nuclear-burst} slices, due to the more frequent injection of stellar feedback in the \texttt{high-z} model. The resulting higher wind pressure also leads to a higher cool and intermediate phase pressure and density. The \texttt{high-z} model's higher SFR leads to a faster outflow in all phases near the disk. A final obvious difference in the \texttt{nuclear-burst} simulation profiles is a break in all measures of the hot phase at a radius of $\gtrsim8~\textrm{kpc}$, which marks the reverse shock that has already exited the volume in the \texttt{high-z} simulation.

\subsection{Outflow Dust Properties} \label{subsec:dust_properties}

\begin{figure*}[h!]
\centering
\includegraphics[width=0.85\textwidth]{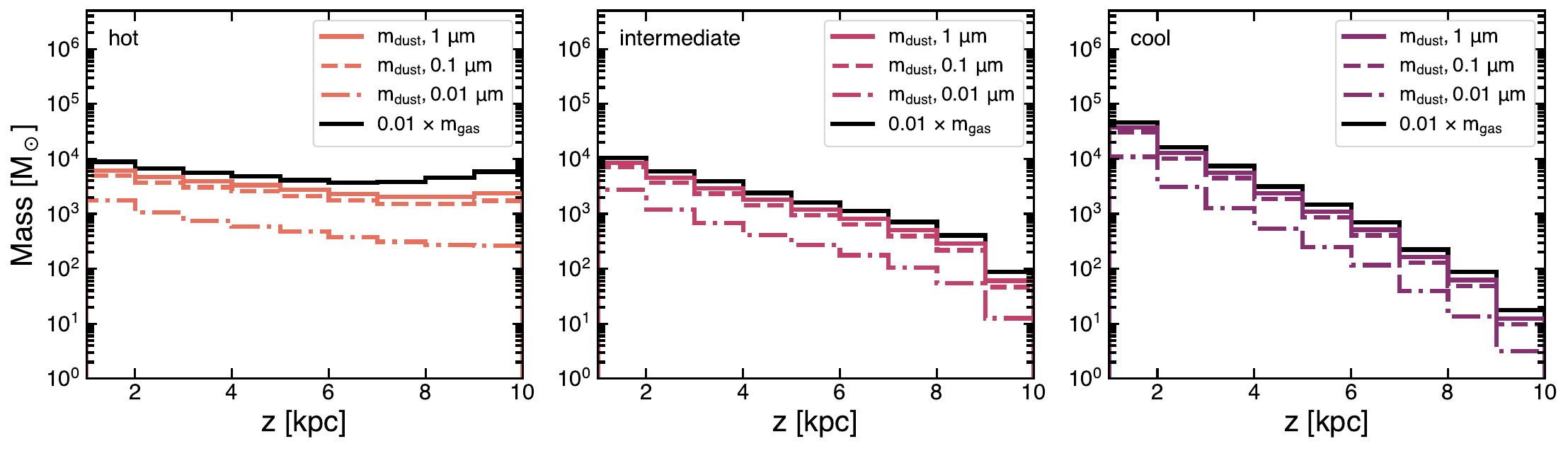}

\vspace{0.2cm}

\includegraphics[width=0.85\textwidth]{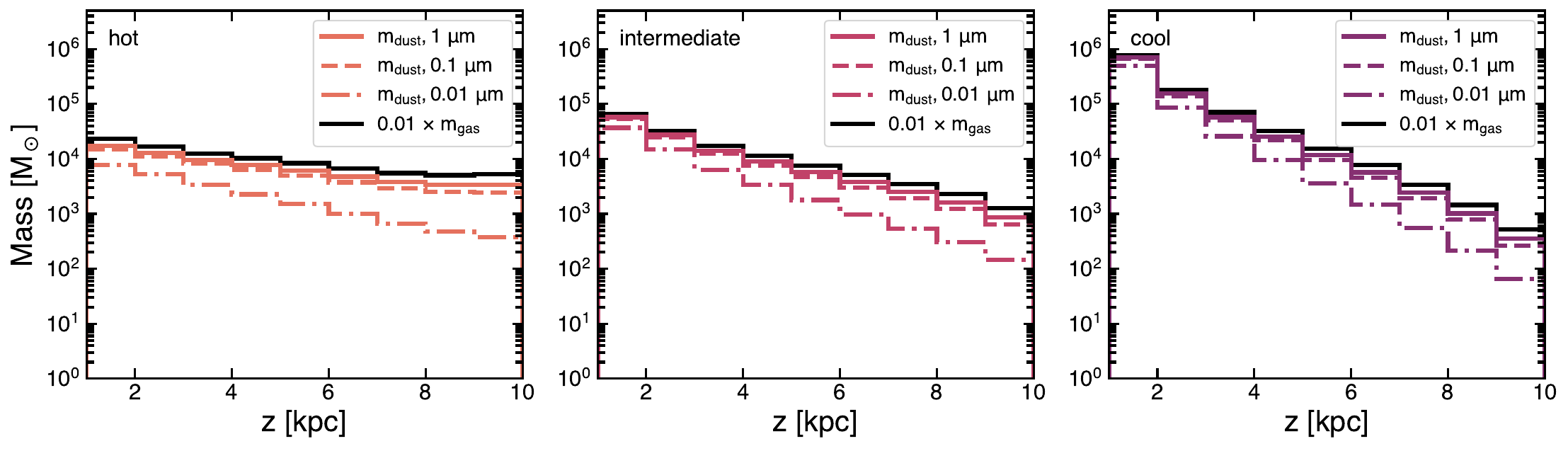}
\caption{Vertical mass profiles of the $30~\textrm{Myr}$ snapshot for the \texttt{nuclear-burst} (top) and \texttt{high-z} (bottom) simulations (excluding the disk region of $z<1~\textrm{kpc}$). Colored lines show dust masses (with the line color indicating the gas phase and the line style indicating the grain radius) and black lines show the gas mass scaled by the initial dust-to-gas ratio. Masses were measured using the vertical bins described in Section~\ref{subsec:dust_properties}.}
\label{fig:dust_profiles}
\end{figure*}

\begin{figure*}[h!]
\centering
\includegraphics[width=\textwidth]{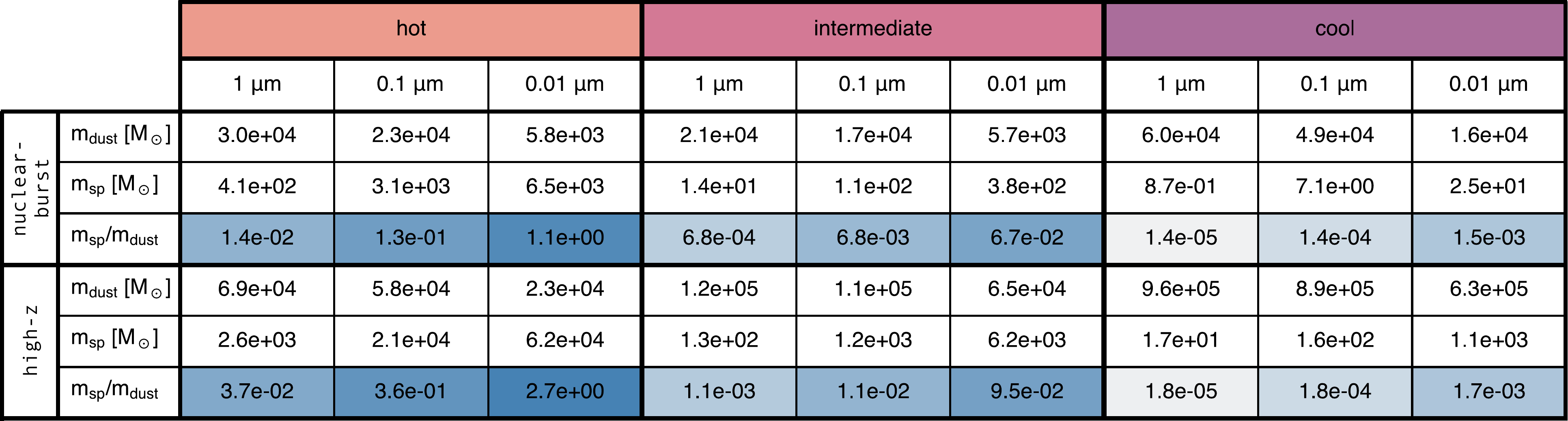}
\caption{Each row shows the dust mass, cumulative sputtered mass, and the ratio of these two quantities summed over the entire $|z|>1~\textrm{kpc}$ volume at 30 Myr. The $m_\mathrm{dust}$ and $m_\mathrm{sp}$ here are equivalent to the binned masses shown in Figures~\ref{fig:dust_profiles} and \ref{fig:sputtering_profiles}, but summed over $z$. Blue shading highlights the sputtering efficiency, with darker cells indicating higher values of $\log_{10} (m_\mathrm{sp}/m_\mathrm{dust})$, compared between all grain sizes, phases, and simulations.}
\label{tab:dust_mass_table}
\end{figure*}

At each time step during the simulations, we record the total dust mass ($m_\mathrm{dust}$), gas mass ($m_\mathrm{gas}$), and sputtered dust mass ($m_\mathrm{sp}$) in three separate gas phases, using the phase definitions described in Section~\ref{subsec:radial_profiles}. We record spatial information for each of these quantities using bins defined by the $z$-coordinate. For $|z|>1~\textrm{kpc}$ (measured from the disk mid-plane), these bins consist of 18 $10\times10\times1~\textrm{kpc}^3$ volumes. For the disk region (i.e. $|z|\leq1~\textrm{kpc}$), these quantities are summed in a single $10\times10\times2~\textrm{kpc}^3$ volume. The resulting vertical dust mass profiles are plotted in Figure~\ref{fig:dust_profiles} for the $|z|>1~\textrm{kpc}$ bins at $30~\textrm{Myr}$ for the \texttt{nuclear-burst} (top) and \texttt{high-z} (bottom) models. We also show the vertical gas mass profiles for each phase, multiplied by the initial cool-phase dust-to-gas ratio, to compare with the dust mass profiles.

From Figure~\ref{fig:dust_profiles}, we can see that substantial dust masses exist in all phases. The cool phase is the dustiest at small radii, followed by the intermediate and hot phases. The cool-phase $m_\mathrm{dust}$ falls with $z$ following the cool-phase $m_\mathrm{gas}$, and declines more rapidly than the $n \sim r^{-2}$ median density distribution for the cool phase in Figure~\ref{fig:radial_profiles} would imply. This is because the cool gas accelerates relatively slowly in the outflow and, thus, is still building up in mass at large $z$ at this point in the simulation. The intermediate phase profiles also drop off at large $z$ since there is little cool gas, but they decline less steeply, following a shape that is intermediate between the hot and cool phase profiles. On the other hand, the hot phase $m_\mathrm{dust}$ remains relatively flat throughout the outflow, in line with the hot phase $m_\mathrm{gas}$. This is because the hot-phase gas is evenly distributed throughout the volume (although there is a slight drop with distance relative to $m_\mathrm{gas}$ due to the effects of sputtering).

Despite the decline in cool phase $m_\mathrm{dust}$ with $z$ exhibited in Figure~\ref{fig:dust_profiles}, the cool phase still dominates the total outflowing dust mass in both simulations. This can be seen in Table~\ref{tab:dust_mass_table}, which shows the integrated dust mass for the entire $|z| > 1$ kpc volume. In both simulations, the values of the cool phase total $m_\mathrm{dust}$ are consistently higher than for hot and intermediate phases, even for the smaller grains. This is consistent with the typical picture for gas in outflows, with the cool phase being the mass-dominating phase. In addition, the cool phase is the only phase that initially contains dust.

Given this, the hot- and intermediate-phase dust masses are still higher than one might naively expect---especially for the hot phase, where $t_\mathrm{sp}$ is the shortest. Even the $0.01~\mu\textrm{m}$ grains are abundant in the hot and intermediate phases, despite being $100\times$ more susceptible to sputtering in these environments. For example, for $0.1~{\mu\textrm{m}}$ grains at $r=0~\textrm{kpc}$, the wind sputtering times based on the median gas properties are roughly 5 and $6~\textrm{Myr}$ for \texttt{nuclear-burst} and \texttt{high-z}, respectively. These timescales are short compared to the outflow advection time, which can be approximated as $t_\mathrm{adv}=z/v_w$. For an outflow height of $z=10~\textrm{kpc}$ and a wind velocity of roughly $v_w=800~\textrm{km}\,\textrm{s}^{-1}$, $t_\mathrm{adv}\sim12~\textrm{Myr}$.

\begin{figure}
\centering
\includegraphics[width=0.45\textwidth]{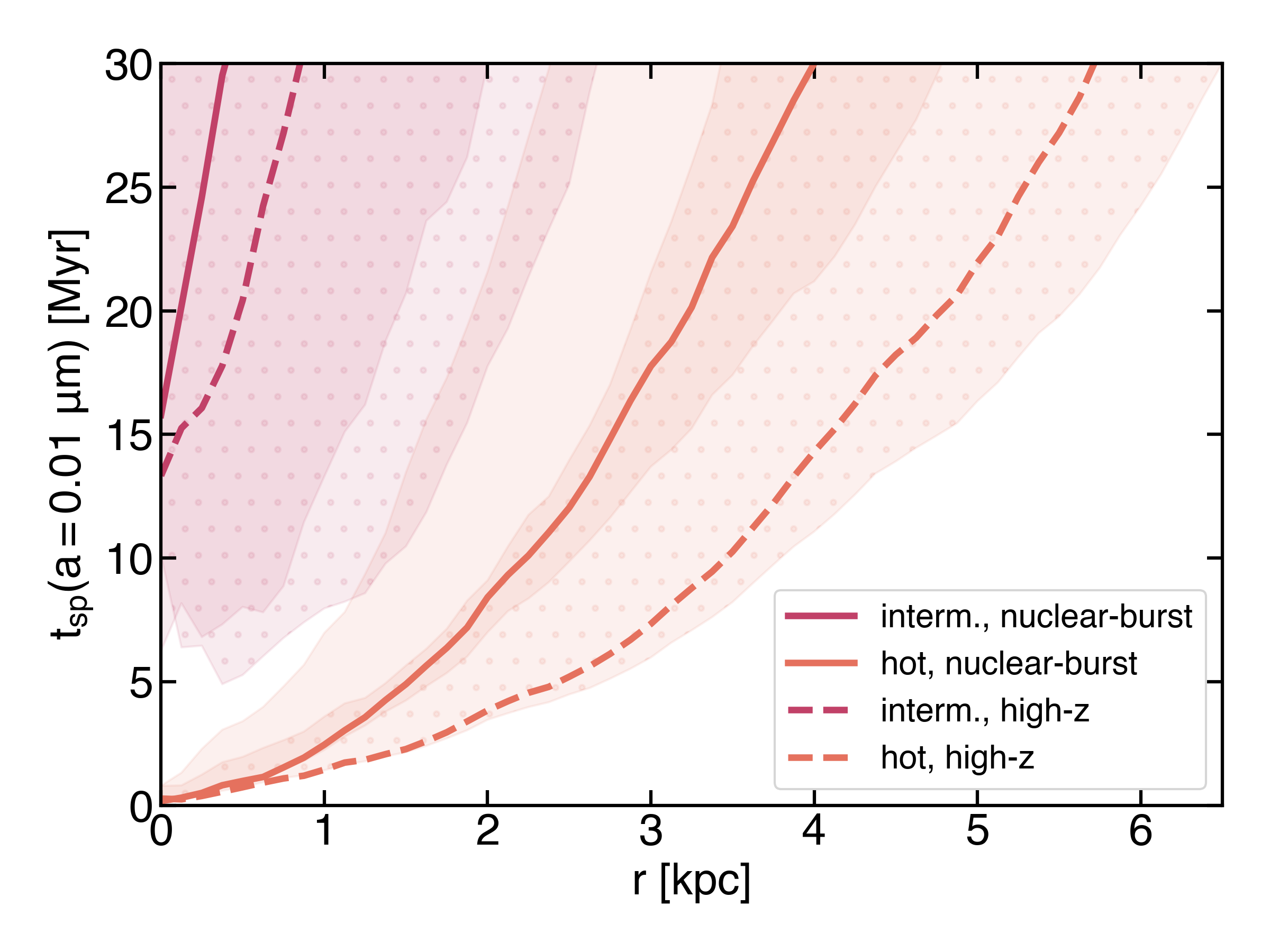}
\caption{Sputtering times ($t_\mathrm{sp}$, Eq.~\ref{eq:sput-timescale}) for $0.01~{\mu\textrm{m}}$ grains in the \texttt{nuclear-burst} (solid lines) and \texttt{high-z} (dashed lines) outflows. We use average values of the radial profiles at $30~\textrm{Myr}$ (similar to those shown in Figure~\ref{fig:radial_profiles}, but weighted by volume instead of density) to calculate $t_\mathrm{sp}$. The extent of the $y$-axis is $30~\textrm{Myr}$, the time of the simulation snapshot we focus on. The shaded regions show the range of $t_\mathrm{sp}$, calculated using the values of $25-75$ percentile range of the radial profiles. The solid region corresponds to the \texttt{nuclear-burst} simulation, and the dotted region corresponds to \texttt{high-z}.}
\label{fig:t_sp_profiles}
\end{figure}

Although the dynamical time is thus longer than the $r=0~\textrm{kpc}$ sputtering times, $t_\mathrm{sp}$ decreases by orders of magnitude with increasing distance from the disk due to the drop in $n_\mathrm{hot}$ and $T_\mathrm{hot}$ with $r$ shown in Figure~\ref{fig:radial_profiles}. Figure~\ref{fig:t_sp_profiles} illustrates the effect of this rapid decline on the sputtering time, where $t_\mathrm{sp}(a=0.01~{\mu\textrm{m}})$ is plotted as a function of $r$ for the hot and intermediate phases (calculated using average radial profiles similar to those shown in Figure~\ref{fig:radial_profiles}, but volume-weighted). We also use the shaded regions in Figure~\ref{fig:t_sp_profiles} to show the range of $t_\mathrm{sp}$ covered by the $25-75$ percentile values of the radial profiles. Here, the $y$-axis ranges from zero to the simulation runtime, highlighting the decreasing relevance of sputtering as the wind properties become less extreme. Since dust in the hot phase of these simulations moves at the wind speed, it does not take long for it to reach regions of the wind where $t_\mathrm{sp}>>t_\mathrm{adv}$. Therefore, we find that \textit{large populations of dust can exist in the hot phase}.\footnote{We note that these simulations assume perfect dynamical coupling between gas and dust, so we may underestimate the dust entrainment time in the hot phase, potentially leading to overestimates of dust survival (especially for larger grains).}

Now, we compare the dust masses between grain sizes in Figure~\ref{fig:dust_profiles}. Generally, $1~{\mu\textrm{m}}$ grains have the highest masses across phases, followed by the two smaller grain sizes, as expected due to the $a$-dependence in Eq.~\ref{eq:sput-timescale}.\footnote{Note that the offset in $m_\mathrm{dust}$ between the different grain sizes in Figure~\ref{fig:dust_profiles} is somewhat ``artificially" inflated. Since the disk central gas densities are quite high ($n\sim200~\textrm{cm}^{-3}$), some sputtering occurs in the initial dust distribution, lowering the cool phase dust-to-gas ratio in the disk and decreasing the amount of dust available to launch into the outflow. This effect is only significant for $0.01~{\mu\textrm{m}}$ grains, and has a larger impact on the \texttt{nuclear-burst} simulation since its outflow is primarily generated from the central disk region. If we correct the profiles in Figure~\ref{fig:dust_profiles} for this effect by normalizing each grain size by the total dust mass in the disk, we find that the offset in $m_\mathrm{dust}$ between grain sizes decreases slightly, bringing the $z=1~\textrm{kpc}$ values of $m_\mathrm{dust}$ closer to one another. However, this offset in the $z=1~\textrm{kpc}$ $m_\mathrm{dust}$ bin still exists to some degree in all cases (except for the cool phase in the \texttt{high-z} simulation), suggesting that sputtering occurs as dust is being entrained in the outflow in the $z<1~\textrm{kpc}$ bin.} In the \texttt{nuclear-burst} simulation, the masses of all grain sizes in the cool and intermediate phases stay relatively constant relative to each other. In the hot phase, $m_\mathrm{dust}$ of $0.01~{\mu\textrm{m}}$ grains decreases more rapidly with $z$ than for the larger two grain sizes, a result of sputtering. The hot phase $m_\mathrm{dust}$ behaves similarly in the \texttt{high-z} simulation, but the mass of $0.01~{\mu\textrm{m}}$ grains also declines more rapidly with $z$ in the intermediate and cool phases. This demonstrates that sputtering is efficient for $0.01~{\mu\textrm{m}}$ grains even in the cool and intermediate phases, due to the \texttt{high-z} galaxy's higher density outflow.

Comparing the two simulations, the total dust masses in the \texttt{nuclear-burst} model shown in Table~\ref{tab:dust_mass_table} for the hot and cool phases are comparable to one another (except for the smallest grain size, which is significantly affected by sputtering in the hot phase). The \texttt{high-z} simulation, on the other hand, has about an order of magnitude higher $m_\mathrm{dust}$ in the cool phase than in the hot phase. Comparing the two simulations, the \texttt{high-z} cool phase has a dust mass that is about 16, 18, and 38 times higher for $1$, $0.1$, and $0.01~{\mu\mathrm{m}}$ grains than \texttt{nuclear-burst}. This is a direct result of the \texttt{high-z} simulation's higher SFR, which results in a total cool gas outflow mass that is roughly 15 times higher than in the \texttt{nuclear-burst} simulation. As such, we would expect the \texttt{high-z} simulation to be roughly 15 times dustier than \texttt{nuclear-burst}. The relatively high factor of 38 for $0.01~{\mu\textrm{m}}$ grains is a result of efficient environmental shielding from sputtering, which we discuss further in Section~\ref{subsec:sputtering}.

\subsection{Sputtering} \label{subsec:sputtering}

\begin{figure*}
\centering
\includegraphics[width=0.85\textwidth]{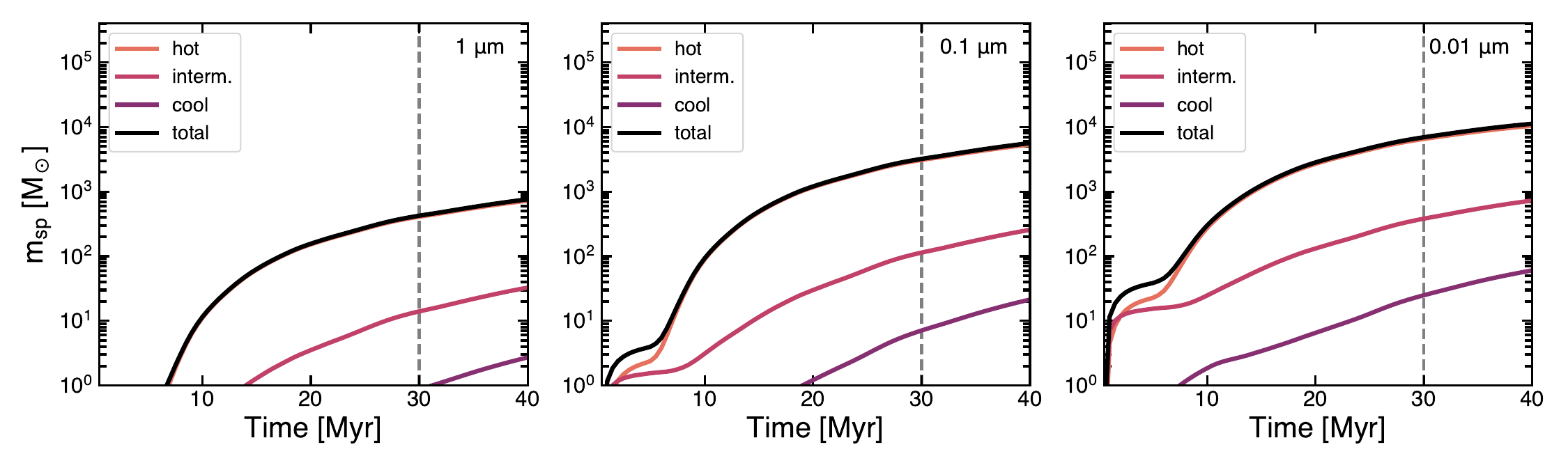}

\vspace{0.2cm}

\includegraphics[width=0.85\textwidth]{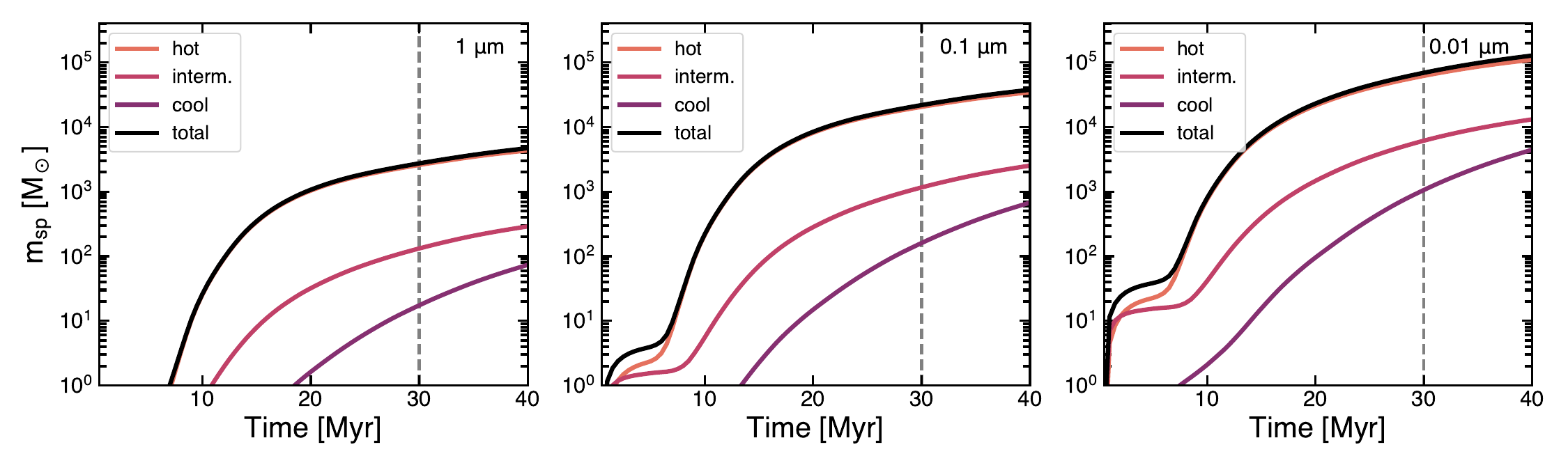}
\caption{The cumulative mass of sputtered dust as a function of time for the \texttt{nuclear-burst} (top) and \texttt{high-z} (bottom) simulations. The colored lines show the total sputtered mass in regions $|z|>1~\textrm{kpc}$ from the disk mid-plane in the hot (orange), intermediate (pink), and cool (purple) phases. The black line shows the sum of $m_\mathrm{sp}$ between all phases. The vertical dashed line marks $30~\textrm{Myr}$, the time of the snapshot we focus on in this section.}
\label{fig:m_sp_evolution}
\end{figure*}

To illustrate how dust evolves due to sputtering in the outflow, in Figure~\ref{fig:m_sp_evolution} we plot the cumulative sum of all sputtered mass in the outflow (for regions $|z|>~1$ kpc) as a function of time. The top panel shows the \texttt{nuclear-burst} simulation, and the bottom panel shows the \texttt{high-z} simulation. Each column shows the values for 1, 0.1, and $0.01~{\mu\textrm{m}}$ grains, with orange, pink, and purple lines representing $m_\mathrm{sp}$ for the hot, intermediate, and cool phases, and the black lines showing the total $m_\mathrm{sp}$ between all phases.

Throughout both simulations, the hot phase accounts for a majority of the measured $m_\mathrm{sp}$, followed by the intermediate phase. The \texttt{nuclear-burst} simulation exhibits little to no sputtering in the cool phase, while the \texttt{high-z} simulation exhibits at least some cool-phase sputtering for all grain sizes. Overall, the \texttt{high-z} model sputters more dust due to its hotter, denser wind and denser cool phase. $0.01~{\mu\textrm{m}}$ grains have the highest sputtered masses in both simulations, reaching a total of $10^4~\textrm{M}_\odot$ and $10^5~\textrm{M}_\odot$ at $40~\textrm{Myr}$ for \texttt{nuclear-burst} and \texttt{high-z}, respectively.

\begin{figure*}
\centering
\includegraphics[width=0.85\textwidth]{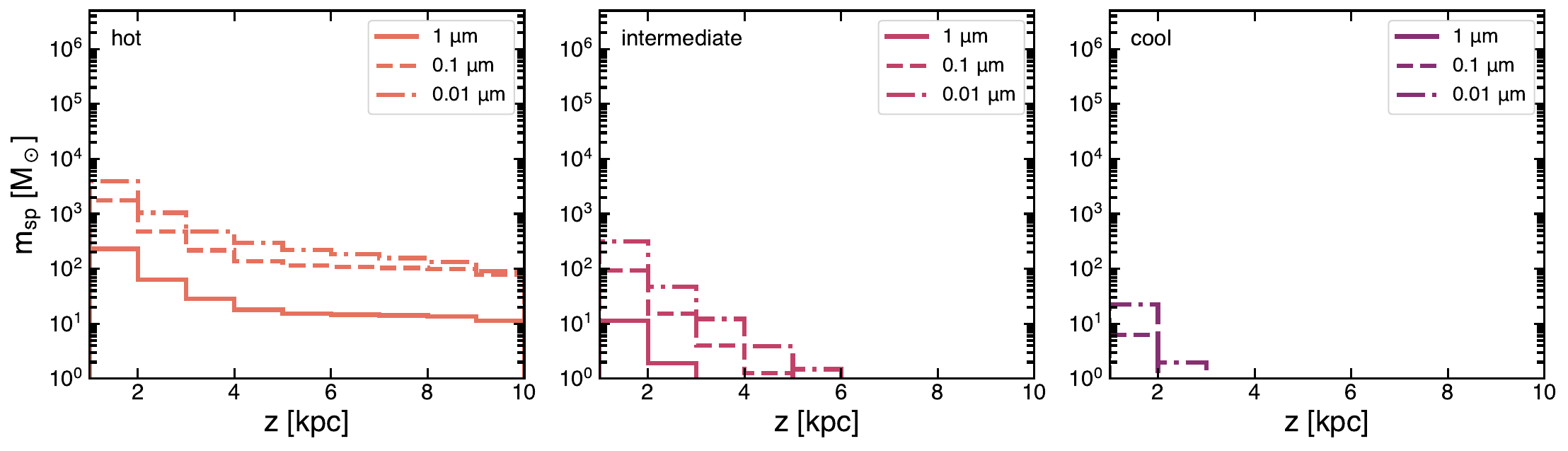}

\vspace{0.2cm}

\includegraphics[width=0.85\textwidth]{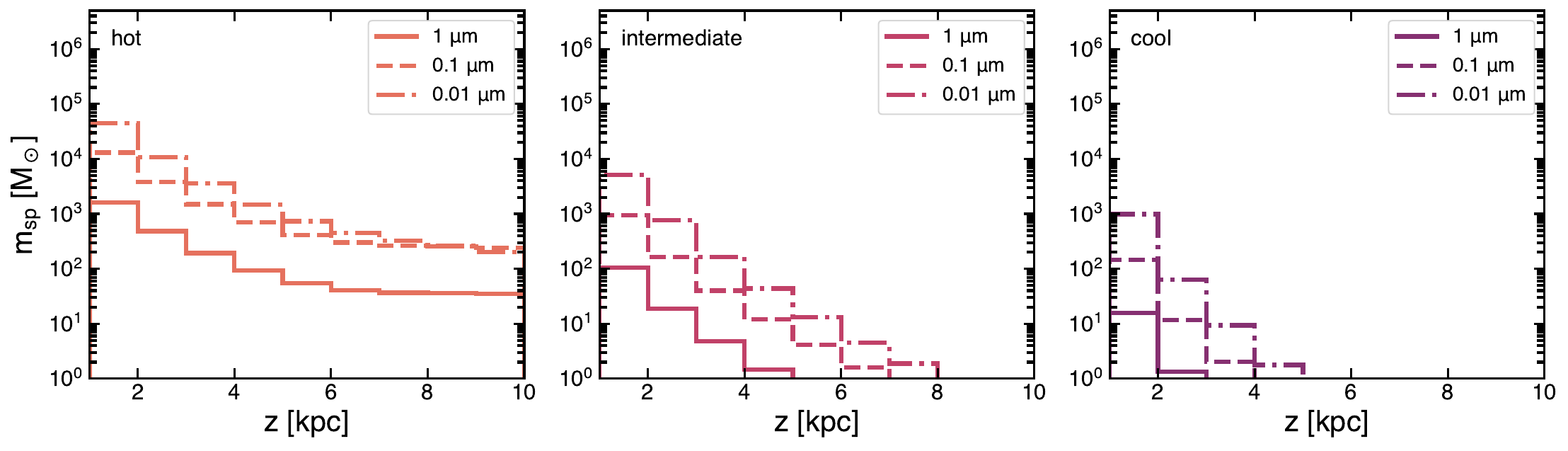}
\caption{Vertical profiles of the cumulative sputtered dust masses in the outflow at $30~\textrm{Myr}$ for the \texttt{nuclear-burst} (top) and \texttt{high-z} (bottom) simulations (measured using the vertical bins described in Section~\ref{subsec:dust_properties}, and excluding the disk region of $z<1~\textrm{kpc}$). The line color indicates the corresponding gas phase and the line style indicates the dust grain radius.}
\label{fig:sputtering_profiles}
\end{figure*}

Returning our focus to the $30~\textrm{Myr}$ snapshot, we can investigate where sputtering occurs in the outflow. We measure the cumulative sputtered dust mass, $m_\mathrm{sp}$, using the vertical bins described in Section~\ref{subsec:dust_properties}. Figure~\ref{fig:sputtering_profiles} shows the resulting vertical cumulative sputtered mass profiles for the \texttt{nuclear-burst} (top) and \texttt{high-z} (bottom) models. For all phases, sputtering is most efficient near the disk and decreases significantly with $z$. This behavior follows naturally from the monotonically decreasing shape of the gas density and temperature profiles and subsequent drop in $t_\mathrm{sp}$, as described in Section~\ref{subsec:dust_properties}. Indeed, $m_\mathrm{sp}$ in the hot phase drops for all grain sizes by at least an order of magnitude by $z\sim5~\textrm{kpc}$. For the same reason, the intermediate and cool phase sputtered masses approach zero at $z=5~\textrm{kpc}$. 

Overall, the intermediate and cool phase sputtering is marginal. This is highlighted in Table~\ref{tab:dust_mass_table}, where we show the total cumulative sputtered mass of each grain size and phase in the outflow. We also show the sputtering efficiency, or the cumulative sputtered mass normalized to the total dust mass in the outflow, for each grain size and phase. These are shown in the rows labeled $m_\mathrm{sp}/m_\mathrm{dust}$ with blue colored cells. Darker blue cells indicate higher fractions of $\log_{10}(m_\mathrm{sp}/m_\mathrm{dust})$, compared between both simulations for all grain sizes and phases. Generally, the normalized sputtered masses show that relatively small fractions of dust are sputtered relative to the total amount in the outflow. The only exception to this is for $0.01~{\mu\textrm{m}}$ grains in the hot phase, which have cumulative sputtered masses that are 1.1 (\texttt{nuclear-burst}) and 2.7 (\texttt{high-z}) times higher than the dust mass in the hot phase of the outflow at this point. The next highest values of $m_\mathrm{sp}/m_\mathrm{dust}$ are 0.13 and 0.36 for $0.1~{\mu\textrm{m}}$ grains in the hot phase. $m_\mathrm{sp}/m_\mathrm{dust}$ for all other phases and grain sizes is $<0.1$.

The \texttt{high-z} simulation exhibits significantly higher $m_\mathrm{sp}$ across phases than the \texttt{nuclear-burst} simulation. This may reflect a higher efficiency of sputtering, however, the higher SFR means that the \texttt{high-z} simulation also has more mass in its outflow to sputter. To understand the efficiency of sputtering compared between the two models, we can again consider $m_\mathrm{sp}/m_\mathrm{dust}$. These normalized sputtered masses show that \texttt{high-z} generally exhibits more efficient sputtering than the \texttt{nuclear-burst} simulation. Significant differences in $m_\mathrm{sp}/m_\mathrm{dust}$ exist between the two simulations in the hot and intermediate phases. The \texttt{high-z} simulation exhibits about $40-60\%$ and $150-180\%$ more efficient sputtering in the intermediate and hot phases, respectively. This is because of the relatively high hot-phase densities and temperatures in the \texttt{high-z} simulation. $10-30\%$ differences exist in the cool phase. These effects are more marginal because the cool phases only differ in density between the simulations, as opposed to both density and temperature in the hot phase.

Morphological differences between the two models may also contribute to their differing sputtering efficiencies. In Figure~\ref{fig:gas_slices}, we can see that the \texttt{nuclear-burst} simulation has a narrow biconical region of $T\gtrsim10^7~\textrm{K}$ and $n\gtrsim0.01~\textrm{cm}^{-3}$ gas near the disk, but beyond this area, $n$ and $T$ drop (caused by the expansion of the wind). Since the \texttt{high-z} model has more distributed star formation in its disk, its wind has a wider region of extremely hot, dense gas near the disk.

The overall inefficiency of sputtering in the outflow is explained by Figure~\ref{fig:t_sp_profiles}. As previously described, this figure shows that the hot phase is the dominant source of sputtering throughout the outflow, but becomes almost irrelevant beyond $r\sim6.5~\textrm{kpc}$ from the disk mid-plane. This figure also illustrates that the average gas properties do not fully capture the broader picture of the outflow because there is a wide spread in the density and temperature. For example, the average intermediate-phase profiles imply that sputtering is only important in the inner $\sim1~\textrm{kpc}$ of the outflow, but the sputtering measurements in Figure~\ref{fig:sputtering_profiles} show sputtering out to much further distances. Similarly, the cool-phase profiles are too long to show up in Figure~\ref{fig:t_sp_profiles}, but we do measure some cool-phase sputtering in Figure~\ref{fig:sputtering_profiles}. Thus, while the average gas properties give us a qualitative sense of how dust evolves in the outflow, the entire distribution of gas properties is needed to fully capture dust evolution.

\subsection{Dust-to-Gas Ratios} \label{subsec:dgr}

\begin{figure*}
\centering

\includegraphics[width=0.545\textwidth]{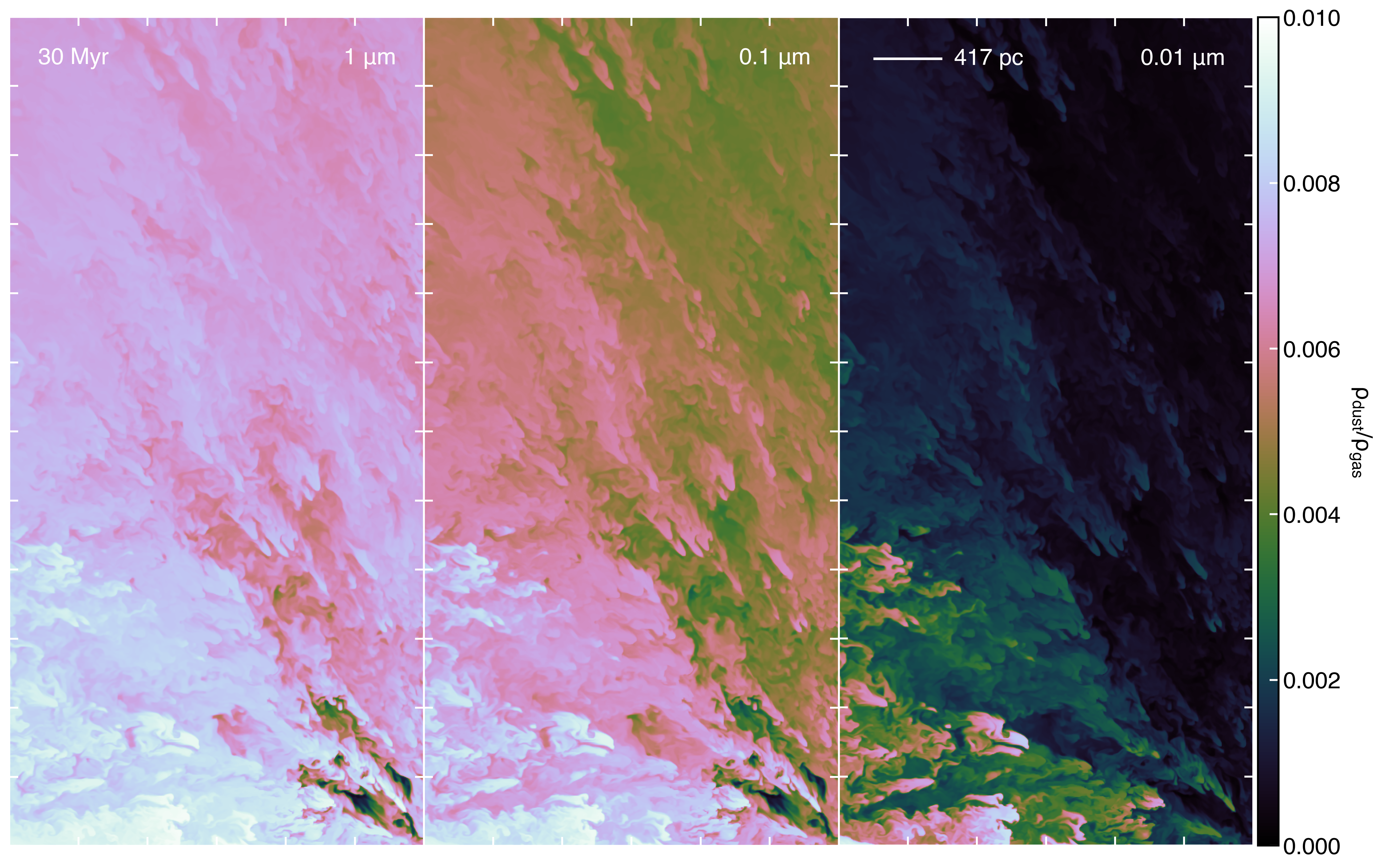}
\includegraphics[width=0.41\textwidth]{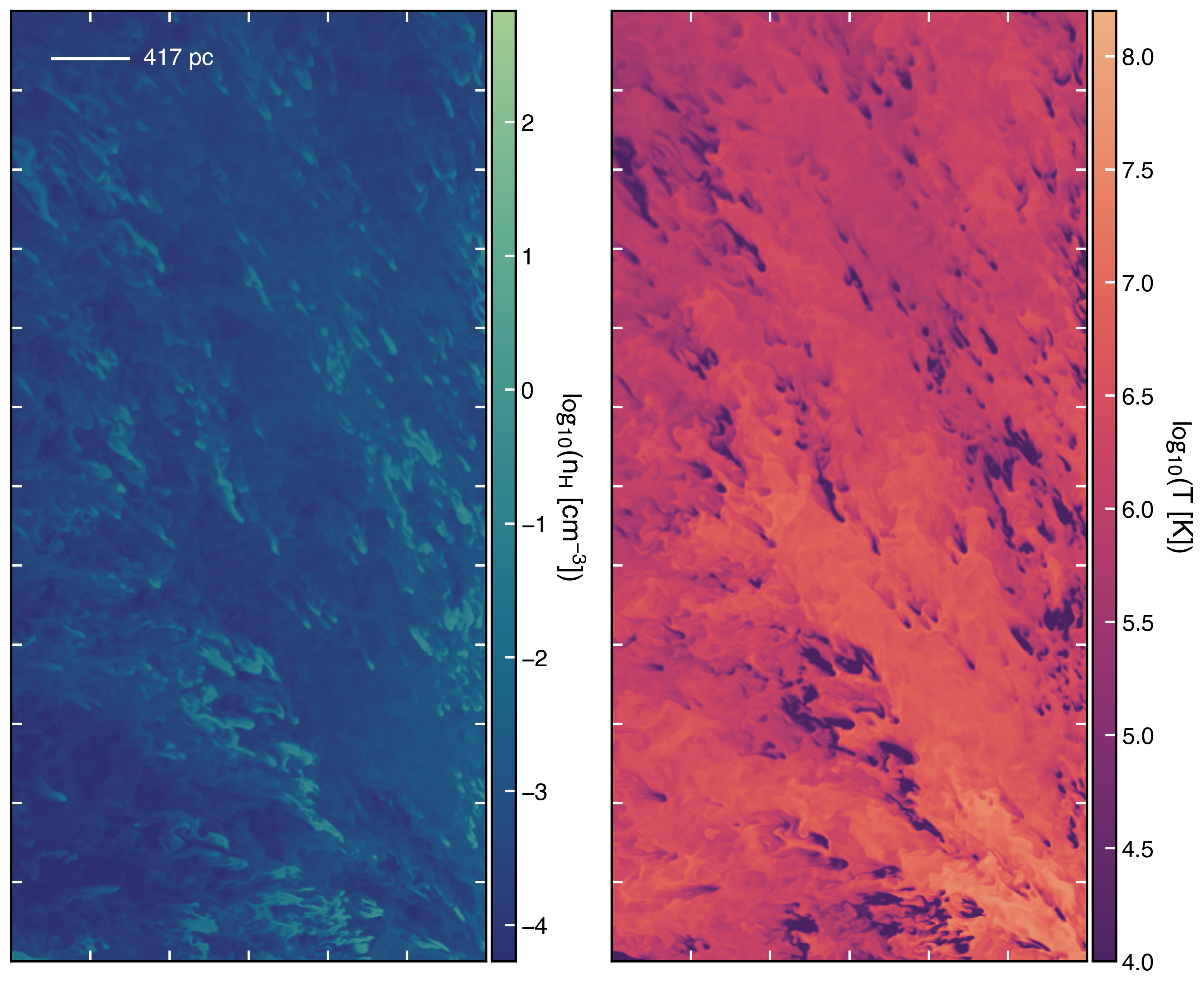}

\vspace{0.5cm}

\includegraphics[width=0.545\textwidth]{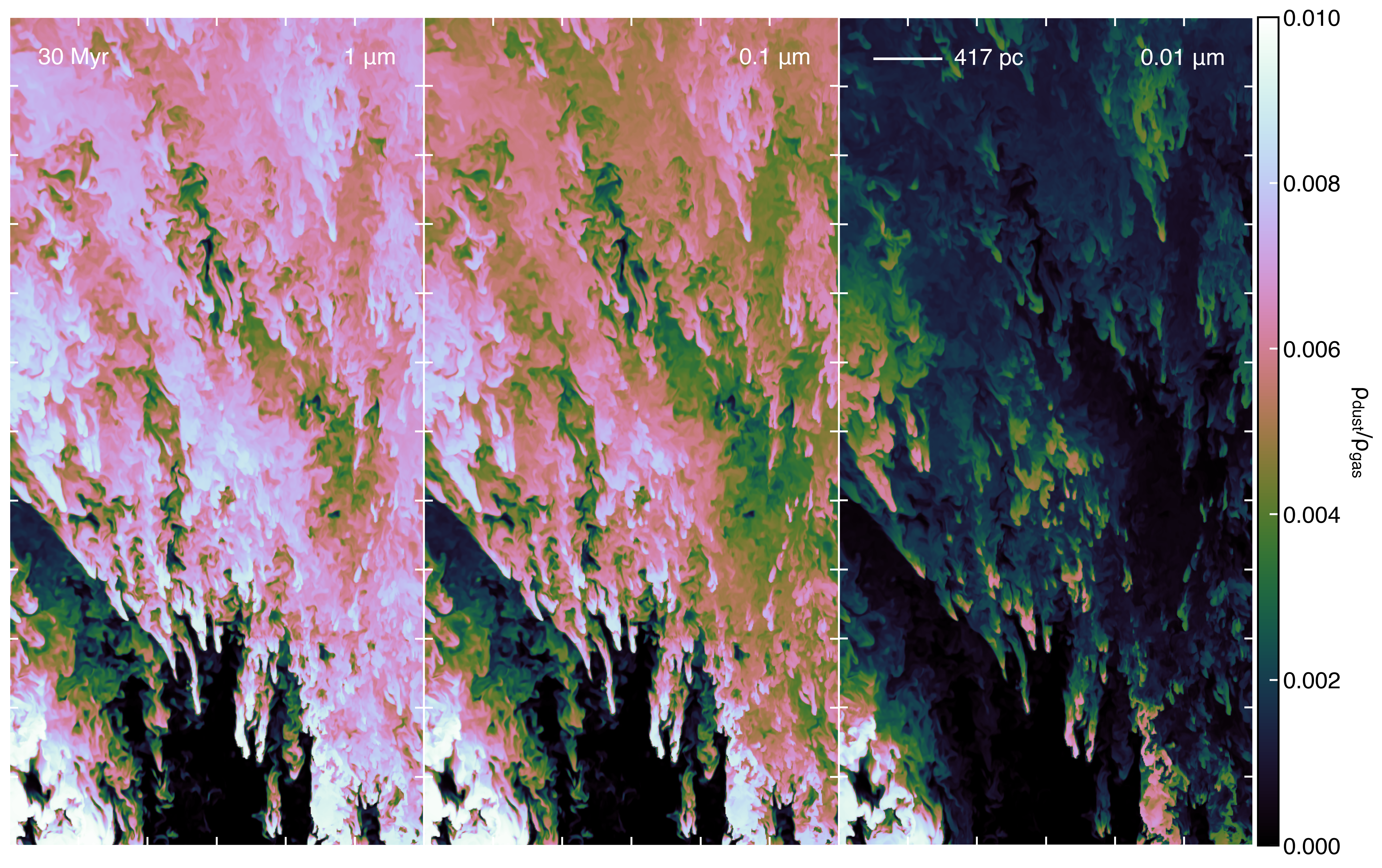}
\includegraphics[width=0.41\textwidth]{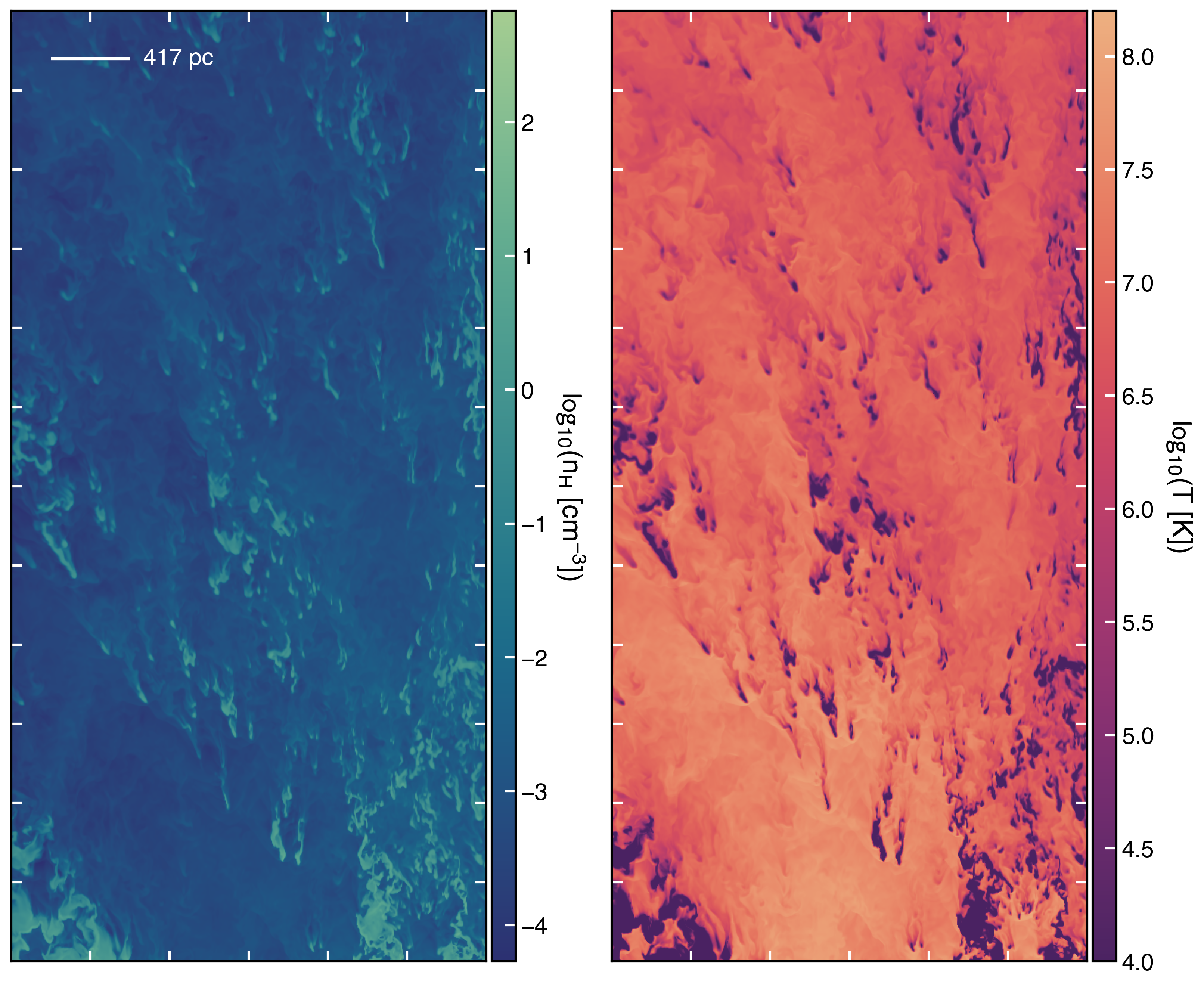}
\caption{Slices of the dust-to-gas ratio for 1, 0.1, and 0.01$~\mu\textrm{m}$ grains (left) and gas density and temperature (right) along the center of the $xz$-plane for the \texttt{nuclear-burst} (top) and \texttt{high-z} simulations. These slices show a zoomed-in region with dimensions $2.4\times4.8~\textrm{kpc}^2$, directly above the disk mid-plane, with the outflow flowing from the bottom edge of the slice.}
\label{fig:dtg}
\end{figure*}

In Figure~\ref{fig:dtg}, we show zoomed-in slices of the dust-to-gas ratio (DGR) along the central $xz$-plane of the \texttt{nuclear-burst} (top) and \texttt{high-z} (bottom) simulations. The left panels show the DGR of 1, 0.1, and $0.01~{\mu\textrm{m}}$ grains, and the right panels show slices of the gas density and temperature in the same region. All panels display the same outflow region, which is located directly above the disk mid-plane, with the outflow flowing upward from the bottom edge. The \texttt{nuclear-burst} outflow originates primarily from the center of the disk (located beyond the right edge of the slice). This results in the angled orientation of the gas flow evident in the \texttt{nuclear-burst} model, whereas the \texttt{high-z} outflow is more perpendicular to the disk.

In both simulations, bubbles of $\sim$zero DGR are visible along the bottom edge of the slices, originating from dust-free supernova ejecta launching out of the disk. These supernova bubbles appear in the gas slices as very hot ($T\sim10^8~\textrm{K}$), low-density regions without cool gas. Clouds of cool, dense gas are accelerated upward by these ejecta, and their cometary tails extend away from the disk. The multi-phase cloud-wind structure is evident in the gas slices, where individual clouds are identifiable as distinct pockets of cool, dense gas surrounded by hotter, and more diffuse intermediate- and hot-phase gas. Intermediate gas fills a higher fraction of the region in the \texttt{nuclear-burst} simulation compared to the \texttt{high-z} simulation.

This multi-phase gas structure is much less obvious in the DGR slices. Although the hot gas is initially dust-free, 1 and $0.1~{\mu\textrm{m}}$ dust grains permeate the hot, intermediate, and cool gas alike. Sputtering is inefficient for these grain sizes, so the cloud-wind mixing that drives cool cloud acceleration also serves to spread dust to the hot and intermediate gas phases without destroying it. This occurs on timescales of roughly the cloud-crushing time, $t_{\rm cc}=(n_{\rm cl}/n_{\rm w})^{1/2}r_{\rm cl}/v_{\rm w}\approx100~{\rm kyr}$ for a cloud overdensity of $10^2$, $r_{\rm cl}=10~{\rm pc}$, and $v_{\rm w}=10^3~{\rm km\,s^{-1}}$ \citep{Klein1994}. Overall, the DGRs of the clouds are the highest, and regions with the hottest gas generally have the lowest DGRs, as these are likely remnants of recent supernova bubbles mixing into the outflow. There is also a mild ($\sim30\%$) gradient in the highest values of DGR with increasing distance from the disk, which happens as the higher cool-phase DGR is diluted through mixing with the dust-free supernova ejecta. The DGR of $0.1~{\mu\textrm{m}}$ grains in both simulations is fairly high throughout the slice, but overall lower than for the largest grain size due to its order of magnitude lower sputtering time.

Hot-phase sputtering times are short for $0.01~{\mu\textrm{m}}$ grains in these regions (as shown in Figure~\ref{fig:t_sp_profiles}), resulting in significantly more obvious cloud-like morphologies for this grain size in Figure~\ref{fig:dtg}. In particular, the classic ``head-tail" structure is visible here, where, on small scales, dusty clouds have extended tails with significantly lower DGRs, and the hottest regions of the outflow are nearly dust-free. This structure is more in line with theoretical predictions from \citet{Richie2024} and \citet{Chen2024} for $0.1~{\mu\textrm{m}}$ grains, where cloud DGRs vary by over an order of magnitude from cloud head to tail as the cloud accretes a mixture of rapidly-cooled pristine hot gas and cool, dusty gas. On an individual cloud level, the variation in DGR for $0.01~{\mu\textrm{m}}$ grains is similar in magnitude to those observed in cloud-wind simulations. The maximum $0.01~{\mu\textrm{m}}$ grain DGR also decreases across the region shown in the slice. This aligns with predictions from \citet{Richie2024}, where variations in DGR happen over scales of $\sim7~\textrm{kpc}$.

\subsection{Dust Outflow Rates} \label{subsec:outflow_rate}

In order to better understand the way that dust populates the CGM, in this subsection we explore the total outflow rates of dust in our simulations. We measure this as the rate at which dust mass flows out of the simulation volume, calculated as 

\begin{equation}
    \dot{m}_\mathrm{dust}=\rho_\mathrm{dust} v_\perp A,
\label{eq:outflow_rate}
\end{equation}

\noindent where $\rho$ is the density of dust in a given cell on the simulation boundary, $v_\perp$ is the gas velocity of the cell in the outward direction of flow from the box, and $A$ is the area of the cell. We sum these values separately for the hot, intermediate, and cool phases across the entire boundary, excluding the disk region ($|z|\leq1~\textrm{kpc}$) and calculating at a cadence of $50~\text{kyr}$. The resulting outflow rates are shown as a function of time in Figure~\ref{fig:dust_outflow_rates}, with each panel showing $\dot{m}_\mathrm{dust}$ for a particular grain size and the colored lines representing the gas phase. Black lines show the total $\dot{m}_\mathrm{dust}$ for each $a$.

The hot wind is the first phase to begin transporting dust out of the volume due to its relatively high speed we define the time at which a given material begins leaving the simulation domain as the time when any of the boundary cells become non-zero in that quantity). It takes $t_\mathrm{adv}\sim 12-14~\textrm{Myr}$ for dust to begin exiting the volume in the \texttt{high-z} and \texttt{nuclear-burst} models. The $1~{\mu\textrm{m}}$ grains are the first to leave the volume, followed by 0.1 and $0.01~{\mu\textrm{m}}$ grains at a delay of $1~\textrm{Myr}$ each in both simulations. This delay is caused by the sputtering of the smaller grain sizes in the initial wind-halo shock. For much of the simulation, the hot phase is responsible for carrying all or almost all of the dust out of the box. This is especially true for the \texttt{nuclear-burst} model, where at $30~\textrm{Myr}$ cool- and intermediate-phase gas and dust have only just begun exiting the box. The intermediate- and cool-phase dust start leaving the box at roughly the same time, which is 17 and $23~\textrm{Myr}$ for \texttt{high-z} and \texttt{nuclear-burst}, respectively. Overall, the \texttt{high-z} model has higher outflow rates and accelerates cool gas faster. When comparing $\dot{m}_\mathrm{dust}$ to the SFR of each model, we see that the \texttt{high-z} simulation is also more efficient at transporting dust out of the galaxy. This is shown by the right axes in Figure~\ref{fig:dust_outflow_rates}, which show $\dot{m}_\mathrm{dust}$ normalized by the galaxy's SFR. Here, we see that the SFR-normalized outflow rate of the \texttt{high-z} model is generally higher than \texttt{nuclear-burst}. Although the \texttt{high-z} galaxy has a higher sputtering efficiency, it also launches about 15 times more cool gas into its outflow than the \texttt{nuclear-burst} galaxy, providing significant additional shielding, which can explain this trend.

\subsection{Dust Survival Fractions} \label{subsec:survival_fracs}

In this subsection, we discuss estimates of the fraction of dust that survives transport in the outflow. We define the dust survival fraction as the quantity

\begin{equation}
    \frac{m_\mathrm{surv}}{m_\mathrm{surv} + m_\mathrm{sp}}.
\end{equation}

\begin{figure*}
\centering
\includegraphics[width=0.9\textwidth]{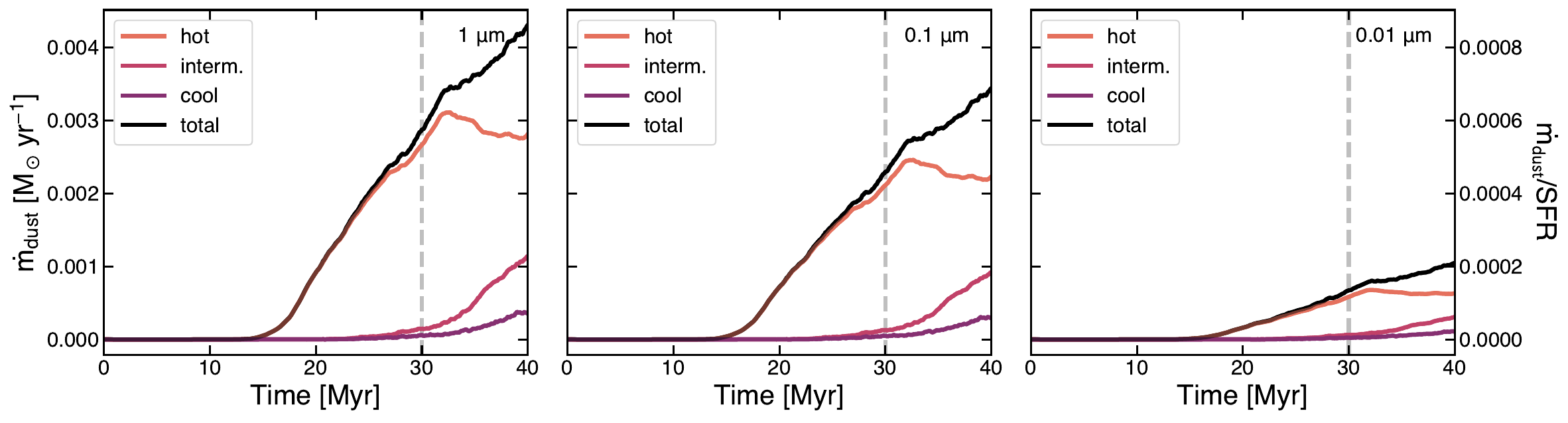}

\vspace{0.2cm}

\includegraphics[width=0.9\textwidth]{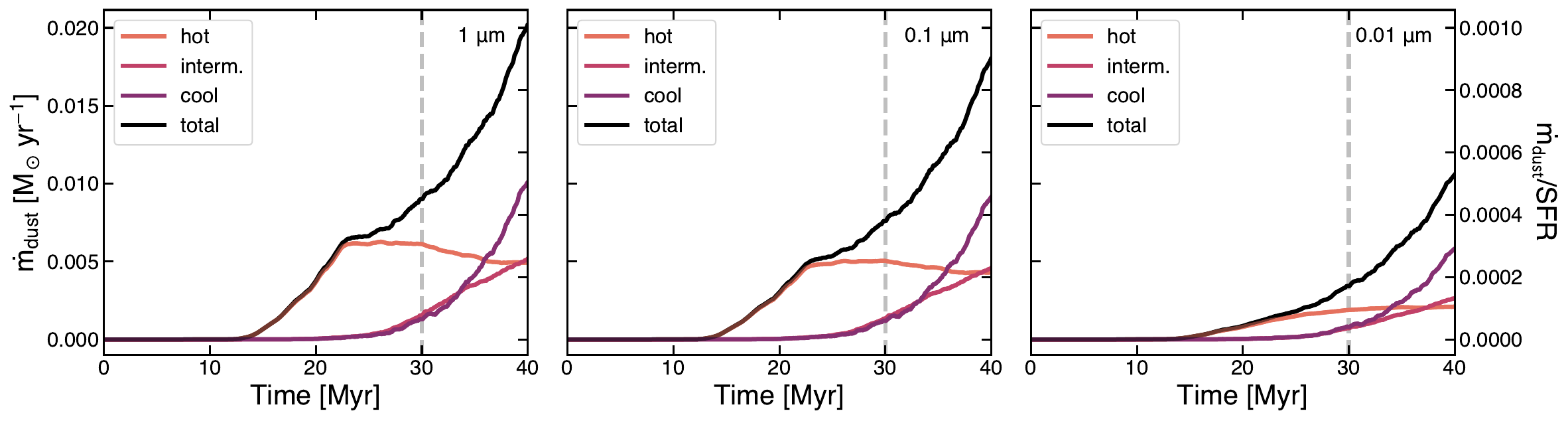}
\caption{Estimates of the rate of dust outflow from the simulation volume, $\dot{m}_\mathrm{dust}$, as a function of time for the \texttt{nuclear-burst} (top) and \texttt{high-z} (bottom) simulations. We estimate $\dot{m}_\mathrm{dust}$ by summing $\rho_\mathrm{dust} v_\perp A$ in all cells along the boundary of the simulation volume every $50~\textrm{kyr}$.}
\label{fig:dust_outflow_rates}
\end{figure*}

\begin{figure*}
\centering
\includegraphics[width=0.85\textwidth]{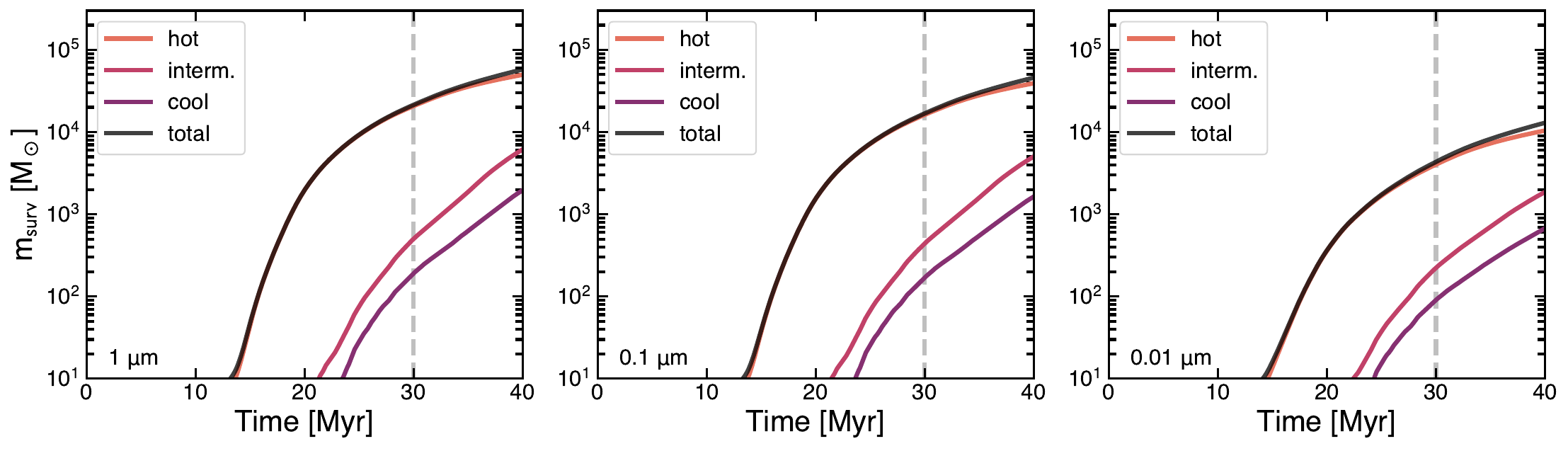}

\vspace{0.2cm}

\includegraphics[width=0.85\textwidth]{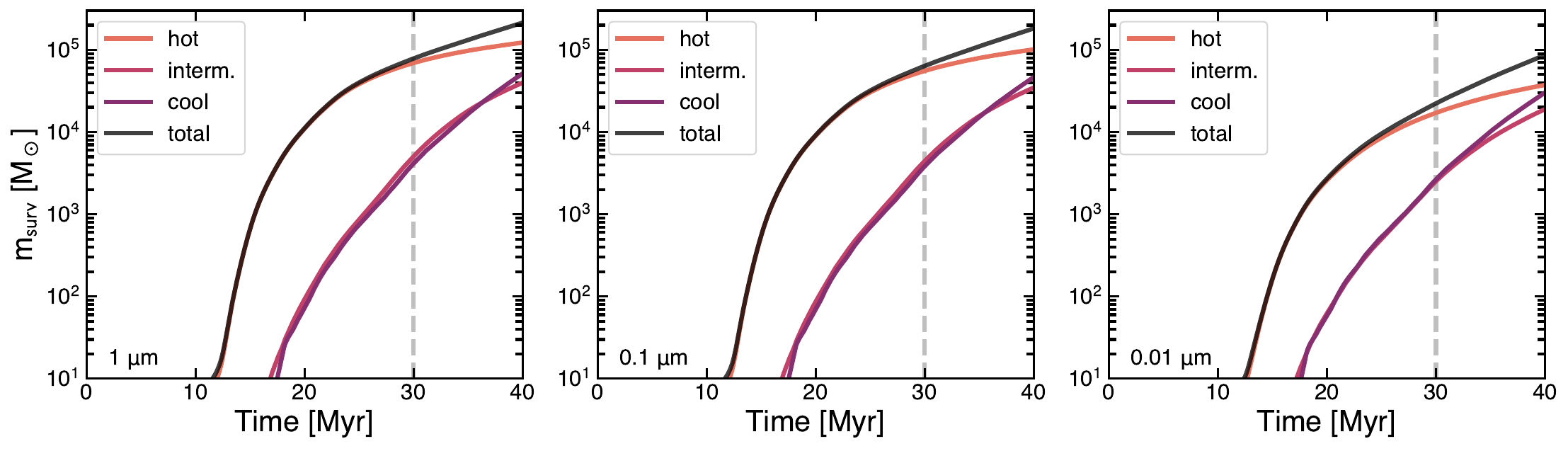}
\caption{Estimates of the cumulative dust mass that has exited the simulation volume as a function of time, or the dust that has ``survived" transport in the outflow. We use the outflow rates shown in Figure~\ref{fig:dust_outflow_rates} to calculate $m_\mathrm{surv}=\sum_{i} \dot{m}_{\mathrm{dust},i}\Delta t$ from $i=0-40~\textrm{Myr}$, where $\Delta t=50~\textrm{kyr}$ is the frequency with which we measure $\dot{m}_\mathrm{dust}$. The colored lines show the total amount of dust that leaves the volume in the hot (orange), intermediate (pink), and cool (purple) phases. The black line shows the sum of $m_\mathrm{surv}$ between all phases. The vertical dashed line marks $30~\textrm{Myr}$, the time of the snapshot we focus on in this section.}
\label{fig:m_surv_evolution}
\end{figure*}

\noindent Here, $m_\mathrm{sp}$ is the cumulative mass of sputtered dust in the outflow for a given grain size, or the value of the black curve in Figure~\ref{fig:m_sp_evolution} at a particular time. The ``survived" dust mass, $m_\mathrm{surv}$, is an estimate of the cumulative dust mass that has exited the box, made using the outflow rates shown in Figure~\ref{fig:dust_outflow_rates}. In particular, we integrate the outflow rates over the simulation time, i.e.

\begin{equation}
    m_\mathrm{surv}=\sum_{i} \dot{m}_{\mathrm{dust},i}\Delta t,
\end{equation}

\noindent where $\Delta t=50~\textrm{kyr}$ is the frequency with which we measure $\dot{m}_\mathrm{dust}$. The resulting cumulative outflow mass estimates as a function of time are shown in Figure~\ref{fig:m_surv_evolution}. 

Overall, we estimate that $98\%$, $84\%$, and $38\%$ of the $1$, $0.1$, and $0.01~{\mu\mathrm{m}}$ dust in the outflow survived in the \texttt{nuclear-burst} model, and $97\%$, $74\%$, and $24\%$ for the \texttt{high-z} model at $30~\textrm{Myr}$. In both models, we measure high fractions of dust survival for all but our smallest grain size, especially compared to theoretical predictions from isolated cloud-wind simulations \citep{Richie2024}.\footnote{Note that because we compare $m_\mathrm{sp}$ and $m_\mathrm{surv}$ at the same simulation time, these fractions probably underestimate dust survival since they do not account for the time it takes for the dust to move to the edge of the box. The non-spherical geometry with which we measure $m_\mathrm{surv}$, however, may counteract this effect, as dust that escapes the side of the box at low $z$ spends less time in the outflow than dust that escapes from other regions of the volume.} We also measure slightly higher dust destruction in the \texttt{high-z} simulation compared to the \texttt{nuclear-burst} simulation due to its relatively hot and dense wind. We discuss these dust survival fractions further in Section~\ref{subsec:non_ideal}.

\subsection{PAH-Sized Grains} \label{subsec:small_dust}

\begin{figure*}
\centering
\includegraphics[width=\textwidth]{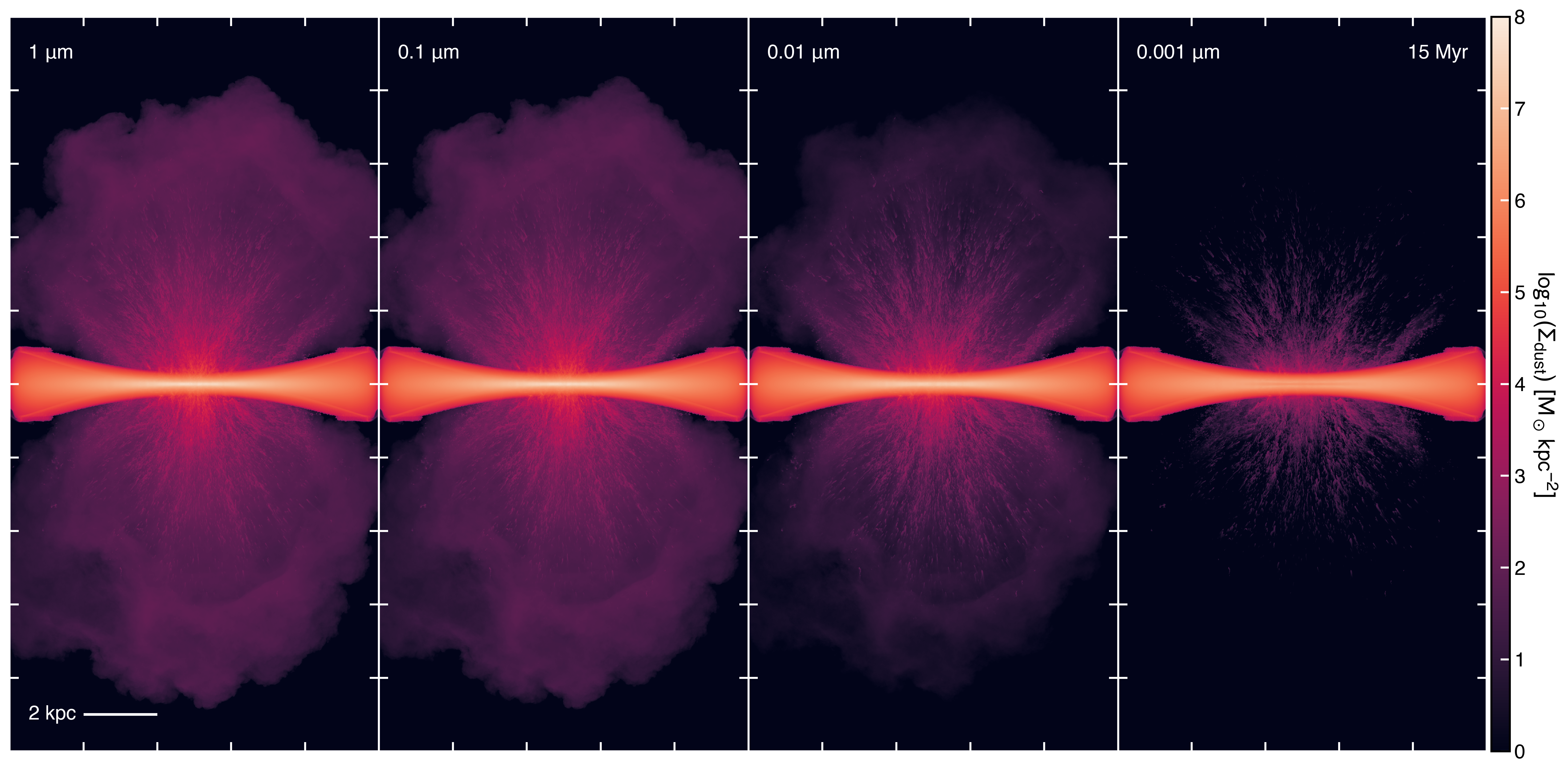}

\vspace{0.5cm}

\includegraphics[width=\textwidth]{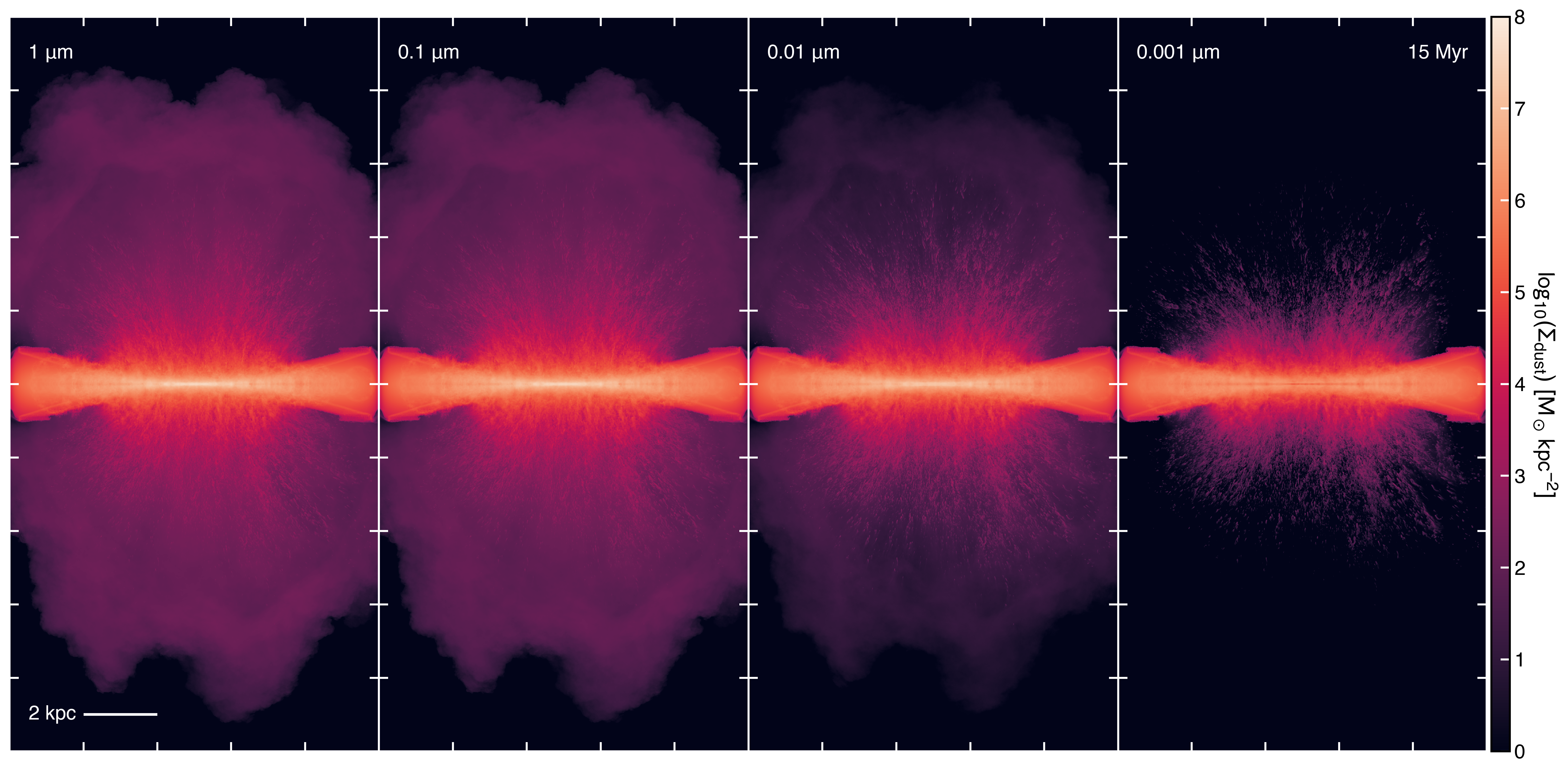}
\caption{Dust surface density projections for the \texttt{nuclear-burst} (top) and \texttt{high-z} galaxy models at $15~\textrm{Myr}$. The dust grain radius is labeled in the upper left corner of each panel. The largest grains in both simulations most clearly show the propagation of the outflow into the initial (dust-free) hot halo. The boundary of the outflow is marked by the contact discontinuity. Some amount of diffuse background dust, which traces the hot wind, can be seen in all but $a=0.001~{\mu\mathrm{m}}$ grains. These grains are strongly sputtered in the hot wind and primarily trace cool, dense gas.}
\label{fig:dust_projection_early}
\end{figure*}

In this subsection, we briefly shift our focus to an earlier simulation snapshot at $15~\textrm{Myr}$ (comparable to the duration of M82's most recent burst period; \citealt{Schreiber2003}). Figure~\ref{fig:dust_projection_early} shows dust density projections of the \texttt{nuclear-burst} (top) and \texttt{high-z} (bottom) simulations at this time for four different grain sizes, including a smaller size of $a=0.001~{\mu\textrm{m}}$. This smallest grain size is similar to the size of PAHs, although PAH destruction is more efficient than spherical dust sputtering \citep{Micelotta2010}. As such, the sputtering of these grains can be thought of as a lower limit for PAH destruction in outflows. We exclude this size in our analysis of later snapshots because significant sputtering of these grains occurs in the densest regions of the disk (see discussion in Section~\ref{subsec:dust_properties}), affecting the outflow dust properties for these grains at later times. At $15~\textrm{Myr}$, enough $0.001~{\mu\textrm{m}}$ grains remain in the disk to discuss their qualitative evolution in the outflow.

In Figure~\ref{fig:dust_projection_early}, the contact discontinuity shows the extent of the outflow formed after $15~\textrm{Myr}$. In both simulations, at least for the largest three grain sizes, the outflow extends to a maximum of $z\sim9~\textrm{kpc}$ from the disk mid-plane. The outflow mostly consists of hot gas beyond $z\gtrsim2~\textrm{kpc}$, as the cool, dense gas that stellar feedback ejects from the disk is still becoming entrained in the wind. The most striking feature of Figure~\ref{fig:dust_projection_early} is the complete lack of $0.001~{\mu\textrm{m}}$ grains beyond $z\gtrsim2~\textrm{kpc}$ from the disk. For $1$ and $0.1~{\mu\textrm{m}}$ grains, significant amounts of dust appear throughout the outflow, strongly tracing the out to the forward shock. The $0.01~{\mu\textrm{m}}$ grains also permeate the outflow, though at lower densities than for the larger two grains. To varying degrees, these grains all trace the cool, intermediate, and hot phases. However, the sputtering timescales of $0.001~{\mu\textrm{m}}$ grains are so short that these grains are entirely absent in all but the cool phase. 

\begin{figure}
\centering
\includegraphics[width=0.45\textwidth]{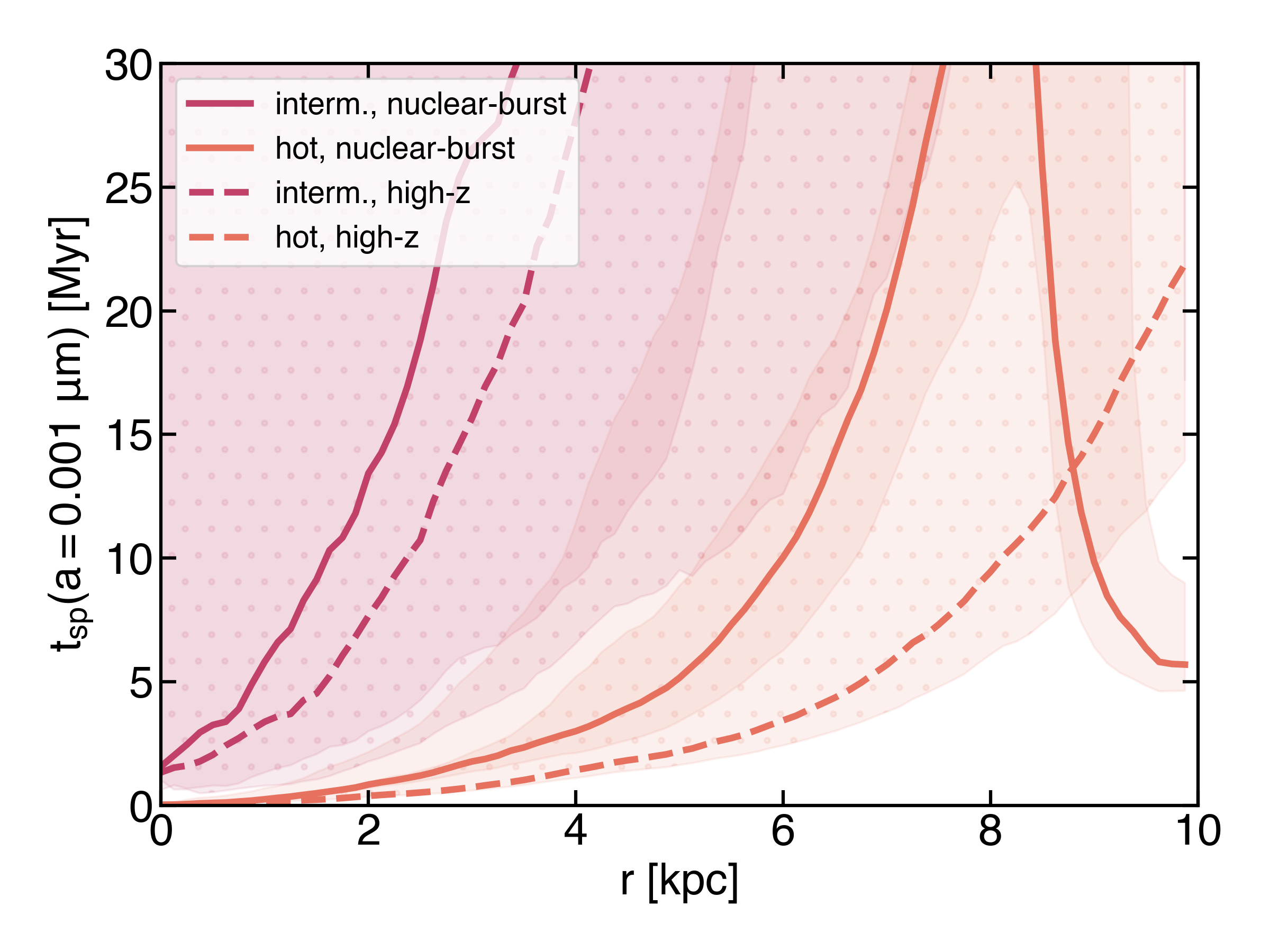}
\caption{Same as Figure~\ref{fig:t_sp_profiles}, but for $0.001~{\mu\textrm{m}}$ grains. The sharp peak in the \texttt{nuclear-burst} hot-phase profile is a result of the reverse shock.}
\label{fig:t_sp_profiles_tiny}
\end{figure}

The absence of these grains is clearly explained in Figure~\ref{fig:t_sp_profiles_tiny}, which shows the same $t_\mathrm{sp}$ radial profiles as Figure~\ref{fig:t_sp_profiles_tiny} at $30~\textrm{Myr}$, but for $0.001~{\mu\textrm{m}}$ grains. These profiles show that the average hot- and intermediate-phase $t_\mathrm{sp}$ for these grains is short compared to the simulation run time for much of the outflow. $t_\mathrm{sp}$ is particularly rapid near the disk, of order $\sim10~\textrm{kyr}$. As a result, the distribution of $0.001~{\mu\textrm{m}}$ grains in the outflow strongly traces the cool phase. Cool clouds in the outflow are extremely well-defined by these grains, with discrete cloud-like structures with cometary tails clearly visible in Figure~\ref{fig:dust_projection_early}. We discuss the morphology of $0.001~{\mu\textrm{m}}$ grains further in Section~\ref{subsec:small_scale_comp}

\section{Discussion} \label{sec:discussion}

\subsection{Comparison With Small-Scale Observations} \label{subsec:small_scale_comp}

Nearby starburst galaxy M82, with its biconical multi-phase outflow and edge-on orientation, provides one of the most detailed views of dust in galactic outflows. M82 has been the subject of many major observing programs, which have left us with a wealth of data on its dusty outflow. Spitzer observations first revealed the existence of a large halo of infrared emission, likely originating from dust and PAHs, that extends out to roughly $6~\textrm{kpc}$ from M82's disk \citep{Engelbracht2006, McCormick2013}. Filamentary substructure is visible in these observations in aromatic feature emission, which is attributed to PAHs. JWST has recently captured the PAH structure at $3.3~{\mu\textrm{m}}$ in M82's outflow in unprecedented detail, with resolution on scales of $\sim1-2~\textrm{pc}$ \citep{Bolatto2024}. Additionally, Hubble observations in the optical, UV, and near-infrared \citep{Mutchler2007} and Chandra X-ray observations \citep{Lopez2020} have provided insight into M82's cool ionized gas, starlight, dust, and hot wind. The expansive wavelength coverage of these observations makes M82's arguably the best-studied galaxy outflow, giving us invaluable insight into the nature of dusty galactic outflows. Many works have focused on using this data to study the phase structure and underlying dust distribution in the M82 outflow, often finding evidence for the association of dust/PAHs and cool ionized gas in outflows, and sometimes also for dust and UV, X-ray, and molecular gas (see, e.g., \citealt{Engelbracht2006, McCormick2013, Leroy2015, Bolatto2024, Fisher2025, Lopez2025, Villanueva2025}).

Despite the detailed nature of our observations of M82, many open questions remain about the structure and composition of gas and dust in M82's outflow. In particular, we have a limited understanding of the underlying distribution of dust and PAH molecules in M82's outflow that result in the emission we observe. From an observational and theoretical perspective, it seems very likely that dust and especially PAHs in outflows are associated with cool gas. However, several scenarios could lead to the filamentary structure of the emission observed in, e.g., \citet{Bolatto2024}. Well-mixed clouds of cool gas, dust, and PAHs could be transported out of the disk by winds, with emission originating from interstellar PAHs on the surfaces of these clouds. PAHs could also be formed inside of these clouds through the shattering of larger dust grains in the cool, dense gas. PAH formation through shattering could also be induced by cloud-wind mixing in the intermediate phase. Or, some combination of these effects could be at play.

While the simulations in this work were not intended to precisely reproduce the structure of the outflow in M82, there are a number of compelling similarities between our results and observations of dust and gas in M82's outflow, which shed light on the underlying dust distribution in M82's outflow. In particular, this work supports the picture that dust is most strongly associated with the cool phase and is distributed on large scales in the outflow. These simulations also support the finding that dust may exist in the intermediate and hot phases for larger grain sizes \citep{Engelbracht2006, McCormick2013}. In particular, our simulated dust distribution is similar to observations from \citet{Engelbracht2006}, which find a continuous distribution of dust emission at longer wavelengths (from warm dust grains), with filamentary structure emerging in aromatic feature emission. We find that very small (PAH-sized) grains do indeed strongly trace the cool phase, and, since they are depleted so strongly in the hot and intermediate phases, they distinguish between phases in a way larger grains do not. This result provides convincing evidence that PAH destruction in the hot phase is probably very efficient, especially given that Eq.~\ref{eq:sput-timescale} likely over-predicts grain lifetimes for small spherical grains, and is roughly $10\times$ higher than the $\sim1~\textrm{kyr}$ lifetime predicted for PAHs in the M82 wind \citep{Micelotta2010}.

We also see similarities between our simulations of $0.001~{\mu\textrm{m}}$ grains and results from \citet{Fisher2025}, which are of particular note since they can be compared to our simulations at a similar spatial resolution ($\sim1-5~\textrm{pc}$). \citet{Fisher2025} carried out an analysis of the detailed distribution of $3.3~{\mu\textrm{m}}$ PAH emission in the inner region of M82's outflow. Their study identified bright individual plumes of $3.3~{\mu\textrm{m}}$ emission extending out of the disk for at least $300-400~\textrm{pc}$. These plumes have widths of order $\sim50-100~\textrm{pc}$ and visibly cloud-like substructure. These clouds appear to have sizes of roughly $\sim10~\textrm{pc}$, some exhibiting cometary tails that extend $\sim30-150~\textrm{pc}$ away from the disk. \citet{Fisher2025} also examined the correlation of $\textrm{Pa}~\alpha$ (tracing cool ionized gas) with $3.3~{\mu\textrm{m}}$ emission, finding a tight correlation between the two. Their study simultaneously found that peaks in PAH emission are anti-correlated with peaks in X-ray emission, possibly suggesting PAH destruction in the hot wind. Some destruction is observed with distance in \citet{McCormick2013}.

\begin{figure}
\centering
\includegraphics[width=0.45\textwidth]{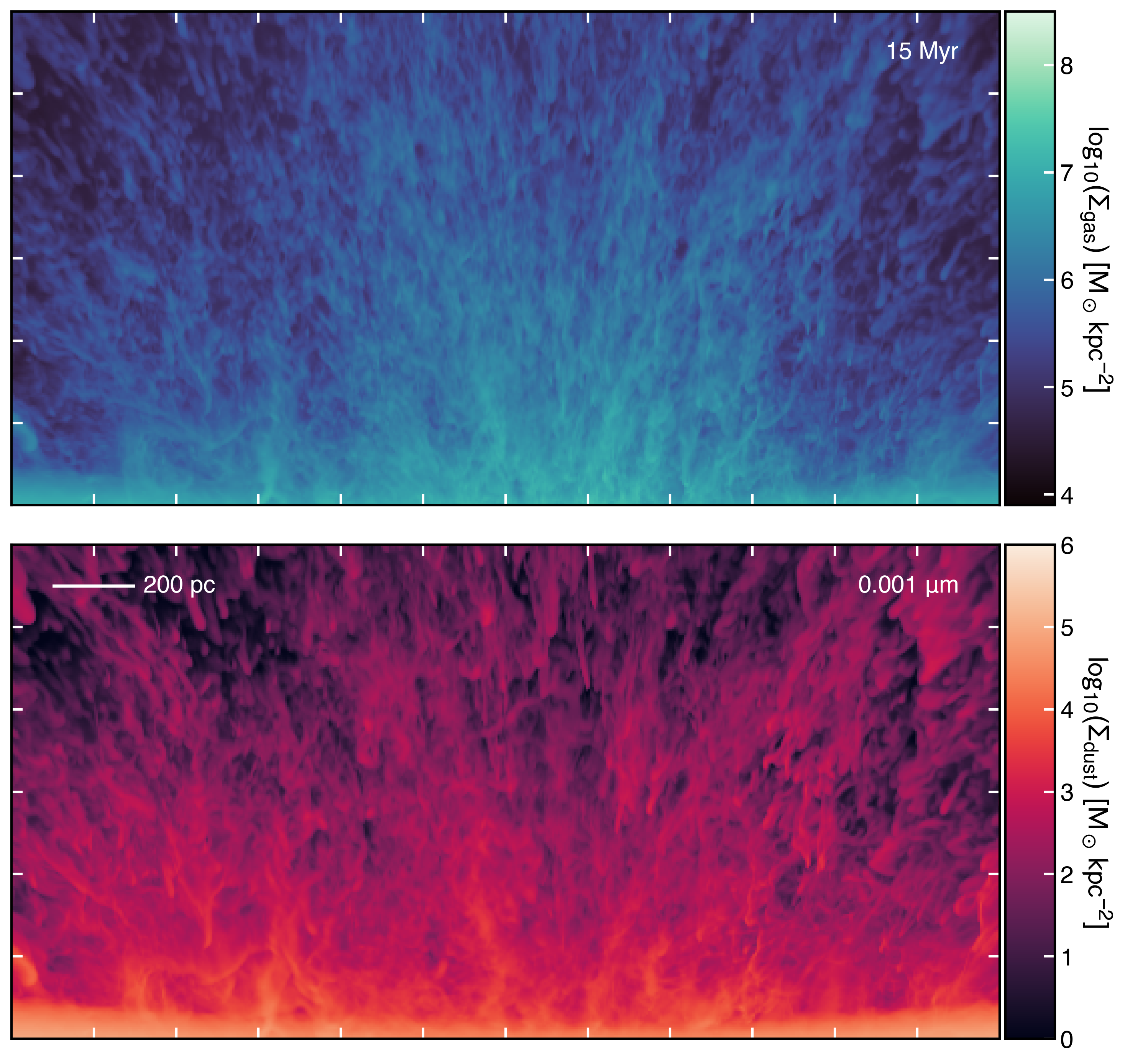} 
\caption{Density projection of gas (top) and $0.001~{\mu\textrm{m}}$ grains (bottom) at the disk-halo interface of the \texttt{nuclear-burst} simulation. Each panel shows a $2.4\times1.2~\textrm{kpc}^2$ sub-region of the volume of the $15~\textrm{Myr}$ snapshot. An animated version of this figure is available in the online article.}
\label{fig:pah_zoom}
\end{figure}

Figure~\ref{fig:pah_zoom} shows zoomed-in regions of the \texttt{nuclear-burst} model analogous to the region covered in \citet{Fisher2025}. Here, we focus on the distribution of gas and $0.001~{\mu\textrm{m}}$ grains at the disk-halo interface, and its connection to the broader outflow. Qualitatively, there is strong agreement between the dust structure seen here and \citet{Fisher2025}. Particularly, similar plume-like structures can be seen extending from the disk in Figure~\ref{fig:pah_zoom}, in both $\Sigma_\mathrm{gas}$ and $\Sigma_\mathrm{dust}$, which bear a striking resemblance to, e.g., Figure 5 of \citet{Fisher2025}. These plumes exist among a somewhat continuous distribution of gas and dust near the disk, and have lengths and widths that are roughly similar to those seen in M82. The densest regions of the dust plumes strongly trace the densest gas, which corresponds to the cool phase. There is no obvious cloud-like substructure to these plumes, although this may be due to our $\sim2.5-5\times$ coarser resolution. Beyond roughly $500~\textrm{pc}$ from the disk, the dominant structure shifts from plume-like and continuous to cloud-like and discrete. This is especially visible in $\Sigma_\mathrm{dust}$, where cometary clouds are prevalent and surrounded by dust-free regions. 

The animation of Figure~\ref{fig:pah_zoom} shows the evolution of the disk-halo region over $20~\textrm{Myr}$, which provides some insight into the possible origin and evolution of the plumes in M82. Overall, the dust plumes in Figure~\ref{fig:pah_zoom} at $15~\textrm{Myr}$ are fairly transient structures. They appear to be the tails of large clouds that are in the process of being launched into the outflow, arising from individual sites of supernova feedback. As these clouds are accelerated, they are shredded into progressively smaller clouds. Beyond roughly $500~\textrm{kpc}$ from the disk, the $\Sigma_\mathrm{dust}$ structure shifts from plume-like and continuous to cloud-like and discrete, and the classic cometary tails emerge. This shift happens as the dust is rapidly destroyed, presumably due to mixing with the hotter phases.

On larger scales, our results suggest that environmental shielding is crucial to the survival of $0.001~\mu\textrm{m}$ grains in outflows. The \texttt{high-z} model can transport an appreciable amount of these grains far from the disk since it has a large amount of cool gas, and thus provides more efficient shielding (see Section~\ref{subsec:outflow_rate}). This is not the case for the \texttt{nuclear-burst} model, which failed to transport any $0.001~\mu\textrm{m}$ grains beyond $r\gtrsim5~\textrm{kpc}$ from the disk. Given this, these results may indicate that shattering plays an important role in producing the populations of PAHs we observe on large scales in M82. Currently, we have a limited understanding of the importance of shattering in outflows. Semi-analytic modeling has shown that shattering in cool gas in the CGM is relevant on timescales of $\sim10^8~\textrm{yr}$ \citep{Hirashita2024}. There is also observational evidence that shattering may be efficient in, e.g., \citet{Katsioli2023}, who observe an increase in the mass fraction of small grains with increasing distance from the disk mid-plane in NGC 891. Our simulations have begun to reveal the underlying dust distribution in outflows and the physics that may be at play in these environments, but more work is needed to fully understand the origin of small extragalactic dust and PAHs at large distances. In future work, we plan to incorporate a shattering model to capture grain-grain collisions to investigate this phenomenon.

\subsection{Comparison With Large-Scale Observations} \label{subsec:large_scale_comp}

Large-scale surveys of reddening in millions of galaxy halos in the local Universe have revealed the ubiquity of dust in the CGM \citep{Menard2010, Peek2015, McCleary2025}. These observations cover large scales, with measurements of dust extinction/surface density, $A_V/\Sigma_\mathrm{dust}$, spanning tens of kpc to tens of Mpc. The shape of these profiles exhibits some dependence on galaxy type. Specifically, studies have found that actively star-forming galaxies have steeper $\Sigma_\mathrm{dust}$ profiles at small radii \citep{Lan2018, Ruoyi2020, Chen2025, McCleary2025}, whereas quiescent galaxies exhibit a lack of dust at small radii and flatter, dustier profiles at large radii \citep{McCleary2025}.

\begin{figure}
\centering
\includegraphics[width=0.5\textwidth]{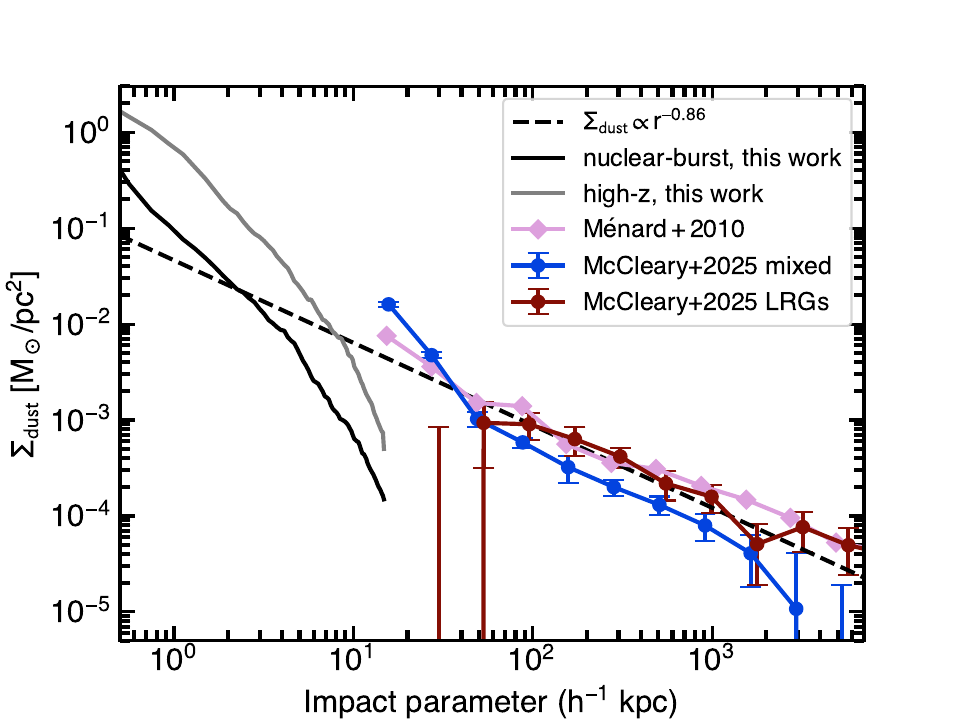}
\caption{Profiles of the average dust surface density, $\Sigma_\mathrm{dust}$, as a function of halo radius. The solid black and gray lines show the average projected $\Sigma_\mathrm{dust}$ of the \texttt{nuclear-burst} and \texttt{high-z} simulations, respectively, at $40~\textrm{Myr}$ (summed over all three grain sizes). The colored lines show the observed halo density profiles from large-scale surveys \citep{Menard2010, McCleary2025}. We show two separate samples of galaxies from \citet{McCleary2025}: redMaGiC luminous red galaxies (LRGs, maroon) and a mixed sample of star-forming and quiescent galaxies from WISExSuperCOSMOS (the WISE $\times$ hi-dens $z>0.5$ sample, blue).}
\label{fig:halo_profile}
\end{figure}

We explore this result in relation to our simulations in Figure~\ref{fig:halo_profile}. Here, the average $\Sigma_\mathrm{dust}$ profiles are plotted for the \citet{Menard2010} and \citet{McCleary2025} samples. The pink diamonds show the \citet{Menard2010} data, and the black dashed line shows a fit to this data, $\Sigma_\mathrm{dust}\propto r^{-0.86}$. Blue and maroon symbols show results from \citet{McCleary2025}, with the light blue showing a mixed sample of quiescent and star-forming galaxies (the WISE$\times$hi-dens $z>0.5$ sample), and the maroon showing luminous red galaxies (LRGs).\footnote{We use $\Sigma_\mathrm{dust}/A_V=0.2~\textrm{M}_\odot\,\textrm{pc}^{-2}\,\textrm{mag}^{-1}$ to convert from $A_V$ to $\Sigma_\mathrm{dust}$, following \citet{McCleary2025}, but we note that assuming a fixed scaling between $\Sigma_{\rm dust}$ and $A_V$ contradicts our finding that the grain size distribution varies significantly in outflows.} At $r<50~\textrm{kpc}$, the dust properties of the mixed and LRG samples differ significantly. Here, the mixed galaxy sample exhibits a steep rise in $\Sigma_\mathrm{dust}$, while the measured $\Sigma_\mathrm{dust}$ for LRGs are too small to appear on this plot (see Figure 3 of \citealt{McCleary2025}). The mixed sample follows a broken power law, with the $r>50~\textrm{kpc}$ slope being roughly in line with the $r^{-0.86}$ behavior of the \citet{Menard2010} data, and at $r<50~\textrm{kpc}$ steepening to $\Sigma_\mathrm{dust}\propto r^{-2.4}$.

We also plot the average projected surface density profiles of the \texttt{nuclear-burst} (black) and \texttt{high-z} (grey) simulations at $40~\textrm{Myr}$ in Figure~\ref{fig:halo_profile}. Like the method used to measure the radial profiles in Section~\ref{subsec:radial_profiles}, we use $30^\circ$ cones opening away from the disk and radial bins of width $\Delta r=0.125~\textrm{kpc}$ (for $r>0.5~\textrm{kpc}$ to exclude the disk) to measure the average $\Sigma_\mathrm{dust}$ for each grain size. The profiles in Figure~\ref{fig:halo_profile} show the summed $\Sigma_\mathrm{dust}$ for all three grain sizes. We use the Planck18 instance of \texttt{astropy.cosmology.FlatLambdaCDM} \citep{PlanckCollaboration2020} to convert from $r$ to impact parameter. We measure slopes of $\Sigma_\mathrm{dust}\propto r^{-2.4}$ and $r^{-2.5}$ for the \texttt{nuclear-burst} and \texttt{high-z} models, respectively, which are very similar to the slope measured for the \citet{McCleary2025} $r<50~\textrm{kpc}$ mixed galaxy sample. Although there is some disagreement in the overall normalization of the profiles, we expect the normalization of these profiles to be affected by our assumptions about galaxy mass, SFR, initial dust-to-gas ratio, etc. The SFR in particular probably plays a significant role in determining the normalization, as evidenced in Figure~\ref{fig:halo_profile} by the difference between the \texttt{high-z} and \texttt{nuclear-burst} profiles.

The agreement between the slopes of our simulations and the \citet{McCleary2025} sample that includes star-forming galaxies supports the picture that star-forming galaxies exist in dusty inner halos resulting from outflows driven by recent star formation. Since these galaxies tend to exist in lower-mass halos, their dusty winds can freely expand over large distances, during which their dust populations may decline through shattering and subsequent sputtering \citep{Hirashita2024}. On the other hand, quiescent galaxies lack this sharp increase in dust density at small radii due to their lack of recent episodes of star formation. However, at larger radii, they may have overall higher density profiles due to their higher stellar masses and correspondingly higher integrated dust output over their lifetime.
 
\subsection{Effects of Global Outflow Conditions on Dust Survival} \label{subsec:non_ideal}

This work predicts high fractions of dust survival for grains $a\geq0.1~{\mu\textrm{m}}$, especially when compared with results from idealized cloud-wind simulations in our previous work. In \citet{Richie2024}, we measured dust evolution due to sputtering in separate regimes of cloud evolution and used those results to estimate the overall fraction of dust survival for a simulated distribution of clouds in an M82-like outflow \citep{Warren2024}. This method resulted in the estimate that $60\%$ of $0.1~{\mu\textrm{m}}$ grains could survive to populate the halo, which is significantly lower than estimates from this work of $84\%$ and $74\%$ for the \texttt{nuclear-burst} and \texttt{high-z} models, respectively.

One of the main reasons for the difference in dust survival between the cloud-wind and full-galaxy simulations is dust shielding in intermediate-phase gas. In the cloud-wind simulations from \citet{Richie2024}, intermediate-phase gas arose only through mixing of the hot wind and cool cloud initiated by hydrodynamical instabilities. Any intermediate-phase gas and dust stripped away from the clouds in these simulations has no option but to gradually mix into the hot wind since there is no additional downstream cloud material for it to interact with. In contrast, in these full galaxy simulations, stripped gas and dust have the opportunity to interact with cool- and intermediate-phase gas from their surrounding environment, providing potential further survival of cool gas and increased shielding of dust.

Another important difference between the isolated cloud-wind simulations in \citet{Richie2024} and this work is the evolution of the hot wind. Cloud-wind simulations typically model the hot background wind using a constant density, temperature, and velocity throughout the box. In \citet{Richie2024}, these wind parameters were $T\sim10^6-10^7~\textrm{K}$ and $n=10^{-2}~\textrm{cm}^{-3}$ throughout the entire $\gtrsim10~\textrm{kpc}$ length of the box. In this work, wind properties are self-consistently generated through supernova feedback injection. As discussed in Section~\ref{subsec:radial_profiles}, the density and temperature of the wind decline considerably as it expands into the outflow. This results in sputtering times at large distances that are orders of magnitude longer than at the disk-halo interface (as shown in Figure~\ref{fig:t_sp_profiles}), the region on which the \citet{Richie2024} wind model is based.

Interestingly, the dust survival fractions in this work consistently increase with time. For example, the \texttt{nuclear-burst} and \texttt{high-z} simulations' survival fractions for $0.1~{\mu\textrm{m}}$ grains increase from 0.84 and 0.74, respectively, at $30~\textrm{Myr}$ to 0.89 and 0.83 at $40~\textrm{Myr}$. That is to say that as the outflow evolves, a higher fraction of dust moves out of the volume than is sputtered. This indicates that sputtering is not as significant a factor in lowering the dust survival fraction as the long cool-phase entrainment times are.

Some differences in the dust survival fractions between cloud-wind and full-galaxy simulations also likely arise from measurement effects. In particular, the definition of the dust survival fraction was slightly different between the two studies in order to limit the measurement to the outflow in this work (see Section~\ref{subsec:survival_fracs} and \citet{Richie2024} for details). The measurement geometries and timescales were also different since material flows out of all sides of the box in the galaxy simulations, as opposed to a single boundary in the cloud-wind simulations.

\subsection{Other Evolution Mechanisms} \label{subsec:other_mechanisms}

In this section, we discuss how the inclusion of more detailed dust physics may affect our results.

A full set of dust evolution mechanisms would include decoupled dust dynamics, shattering, coagulation, accretion, and grain species evolution. There is conflicting evidence as to whether grain growth processes, such as coagulation and accretion, are relevant in these environments \citep{Chokshi1993, Sembach1996, Kirchschlager2020, Priestley2021}. However, the inclusion of grain dynamics is likely to have some effect on our results. Decoupling of dust grains is characterized by the dust drag timescale,

\begin{equation}
    t_\mathrm{drag}\sim\frac{sa}{\rho_\mathrm{g}c_\mathrm{s}},
\label{eq:drag_time}
\end{equation}

\noindent where $s$ is the material density of the grain, $a$ is the grain radius, $\rho_\mathrm{g}$ is the gas density, and $c_s$ is the gas sound speed \citep{Weidenschilling1977}. For a $c_s$ of $15~\textrm{km}\,\textrm{s}^{-1}$ in the cool phase, $\rho_g=5\times10^{-24}~{\textrm{g}\,\textrm{cm}^{-3}}$, $s=3.5~\textrm{g}\,\textrm{cm}^{-3}$, and $a=0.1~\mu\textrm{m}$, the drag time is $\sim15~\textrm{kyr}$. In the hot phase, taking the density to be $\rho_{\rm g}=5\times10^{-27}~{\rm g\,cm^{-3}}$ (roughly the value at $r=1~{\rm kpc}$, where the grains are becoming entrained) and $c_s\sim300~\textrm{km}\,\textrm{s}^{-1}$, the drag time is lengthened to $t_\mathrm{drag}\sim8~{\rm Myr}$.

Since the cool-phase drag timescale is short compared to the relevant dynamical timescales (i.e. the cloud-crushing and advection timescales, $t_{\rm cc}\sim100~{\rm kyr}$ and $t_{\rm adv}=10~{\rm kpc}/v_{\rm w}\sim10~{\rm Myr}$), dust and gas in clouds are likely to remain well-coupled. The drag time in the hot phase is also shorter than, but more comparable to, $t_{\rm adv}$, which could have interesting implications for our results for hot-phase CGM dust loading timescales. The hot-phase $t_{\rm drag}$ is still much shorter than the sputtering time ($t_{\rm sp}\sim22~{\rm Myr}$) in this region, so the overall amount of large dust grains will likely not be reduced by sputtering, but they may advect to the CGM over longer timescales. Hot-phase dust loading occurs through the gradual transfer of dust from cool- to hot-phase gas via mixing, which also imparts momentum to dust grains (see discussion in Appendix D of \citealt{Richie2024}). Consequently, Equation~\ref{eq:drag_time} may overestimate the hot-phase dust entrainment time to some degree. Simulations of outflow dust loading that include dust dynamics are needed to fully understand the degree to which decoupling would decrease the rate of CGM dust loading for large grains. Decoupling can also be significant in supernova shocks \citep{Jones1994}, which has the potential to affect early dust entrainment in the ISM. However, gas and dust in these simulations are initially launched into the outflow through the bulk acceleration of cool ISM material by cluster-scale supernova bubbles, so grains may still remain well-coupled during entrainment. 

As we have seen from the lack of PAH-sized grains in Section~\ref{subsec:small_dust}, shattering likely plays a role in shaping the outflow's grain size distribution to some degree. Over time, the continual shattering of dust may convert large grains into small grains, which are more easily destroyed, leading to the overall decline in outflow dust mass. This is one possible explanation for the systematically lower dust density profiles beyond $\sim50~\textrm{kpc}$ shown in Figure~\ref{fig:halo_profile}. If this is indeed the case, a corresponding increase in gas-phase metallicity in galaxy halos may be observable in galaxy halos.

\section{Conclusions} \label{sec:conclusions}

In this work, we have presented the first large-scale, high-resolution simulations of dust evolution in multi-phase galactic outflows using a scalar-based model for dust sputtering in the Cholla hydrodynamics code. Our main conclusions are as follows:

\begin{itemize}
    \item Galaxies with higher star formation rates and more distributed star formation are more efficient at populating the CGM with dust, due to both their higher mass outflow rates overall, and their higher relative mass of cool gas in the outflow, which results in more efficient shielding of dust in cool gas clouds (Section~\ref{subsec:outflow_rate}).
    \item The cool phase of outflows is the dustiest, but it takes tens of megayears to build up the outflow dust mass in this phase. At early times, the hot phase dominates the mass budget of dust in outflows (Section~\ref{subsec:dust_properties}).
    \item Grains smaller than $\lesssim0.01~{\mu\textrm{m}}$ experience significant sputtering and destruction in gas with $T > 2\times 10^4$ K, which will alter the grain size distribution of dust in the halo relative to the ISM (Section~\ref{subsec:sputtering}).
    \item The hot phase of outflows is a significant carrier of dust that survives transport to the CGM, particularly for large ($\gtrsim0.1~{\mu\textrm{m}}$) grains (Section~\ref{subsec:survival_fracs}).
    \item PAH-sized grains cannot survive at all in high-temperature gas, and require cool gas shielding to travel significant distances in outflows (Section~\ref{subsec:small_dust}).    
    \item Our results show similarities to observations, both in resolved measurements of nearby galaxies and population-wide measurements of distant galaxy halos (Sections~\ref{subsec:small_scale_comp} and \ref{subsec:large_scale_comp}).
\end{itemize}

As a whole, our results provide further evidence that galactic outflows are a very efficient mechanism for driving dust into galaxy halos, potentially explaining the large-scale extragalactic dust populations observed in, e.g., \citet{Menard2010}. These simulations also illustrate that the properties of the dust distribution are likely to evolve during this process, providing new insight into the nature of dust in the circumgalactic medium. This work has shown that average gas properties do not fully predict the effects of global dust evolution, which motivates the need for high-resolution simulations to study this problem. Furthermore, our results indicate that cosmological simulations may underestimate dust outflow rates, because they do not typically resolve the hot phase of outflows. This could explain underpredictions for dust halo masses found in, e.g., \citet{McKinnon2017}. Additional work that incorporates these results into the sub-grid models for dust evolution employed in cosmological simulations is needed to resolve this tension.

\begin{acknowledgments}
We thank the anonymous referee for their constructive comments, which have improved this manuscript. We also
thank Jacqueline McCleary for providing access to the data from \citet{McCleary2025}. This research used resources of the Oak Ridge Leadership Computing Facility, which is a DOE Office of Science User Facility supported under Contract DE- AC05-00OR22725, using Frontier allocation AST181 and SummitPLUS allocation AST200. E.E.S. acknowledges support from NASA ATP grant 80NSSC22K0720 and the David and Lucile Packard Foundation.
\end{acknowledgments}

\software{\texttt{astropy} \citep{2013A&A...558A..33A, 2018AJ....156..123A, 2022ApJ...935..167A}, Cholla \citep{Schneider2015}, \texttt{numpy} \citep{VanDerWalt11}, \texttt{matplotlib} \citep{Hunter07},  \texttt{hdf5} (\href{https://www.hdfgroup.org/}{The HDF Group. Hierarchical Data Format, version 5.}) \texttt{seaborn} \citep{Waskom2021}}

\bibliography{dust}{}
\bibliographystyle{aasjournal}

\end{document}